\documentclass[a4paper,11pt]{article}
\usepackage[a4paper,top=2.5cm,bottom=2.4cm,left=2.4cm,right=3cm]{geometry}
\pdfoutput=1 
\usepackage{jcappub}
\usepackage{bm}
\usepackage{graphicx}
\usepackage{multirow}
\usepackage{amsmath}
\usepackage{amssymb}
\usepackage{hyperref}
\usepackage{color}
\usepackage{xspace}
\usepackage{verbatim}
\usepackage{mathtools}
\usepackage{booktabs}
\usepackage[dvipsnames]{xcolor}
\usepackage[caption=false]{subfig}

\title{\huge Dark Matter Attenuation Effects: Sensitivity~Ceilings~for~Spin-Dependent and Spin-Independent Interactions}

\author[b]{\large QUEST-DMC Collaboration: N. Darvishi,}
\author[c]{J. Smirnov,}
\author[a]{S. Autti,}
\author[d]{L. Bloomfield,}
\author[b]{A. Casey,}
\author[b]{N. Eng,}
\author[d,b]{P. Franchini,}
\author[a]{R. P. Haley,}
\author[b]{P. J. Heikkinen,}
\author[e]{A. Jennings,}
\author[f]{A. Kemp,}
\author[d,b]{E. Leason,}
\author[d]{J. March-Russell,}
\author[a]{A. Mayer,}
\author[d]{J. Monroe,}
\author[a]{D. M{\"u}nstermann,}
\author[a]{M. T. Noble,}
\author[a]{J. R. Prance,}
\author[b]{X. Rojas,}
\author[a]{T. Salmon,}
\author[b]{J. Saunders,}
\author[b,d]{R. Smith,}
\author[a]{M. D. Thompson,}
\author[a]{A. Thomson,}
\author[b]{A. Ting,}
\author[a]{V. Tsepelin,}
\author[b]{S. M. West,}
\author[a]{L. Whitehead,}
\author[a]{D. E. Zmeev}

\affiliation[a]{\textit{Department of Physics, Lancaster University, Lancaster, LA1 4YB, UK}}
\affiliation[b]{\textit{Department of Physics, Royal Holloway, University of London, Egham, Surrey, TW20 0EX, UK}}
\affiliation[c]{\textit{Department of Physics, University of Liverpool, Oxford Street, Liverpool, L69 7ZE, UK}}
\affiliation[d]{\textit{Department of Physics, University of Oxford, Keble Road, Oxford, OX1 3RH, UK}}
\affiliation[e]{\textit{RIKEN Center for Quantum Computing, RIKEN, Wako, 351-0198, Japan}}
\affiliation[f]{\textit{UKRI STFC Rutherford Appleton Laboratory, Particle Physics Department, Harwell, Didcot OX11 0QX, UK}}

\emailAdd{neda.darvishi@rhul.ac.uk}
\emailAdd{juri.smirnov@liverpool.ac.uk}

\bigskip\abstract{\noindent
\noindent Direct detection experiments aimed at uncovering the elusive nature of dark matter (DM) have made significant progress in probing ever lower cross-sections for DM-nucleon interactions. At the same time, an upper limit in the cross-section sensitivity region is present due to DM scattering in the Earth and atmosphere and as a result never reaching the detector.

We investigate the impact of this effect for both spin-dependent and spin-independent interactions.
In contrast to previous studies that assume a straight line path for DM scattering we employ a semi-analytic diffusion model that takes into account the impact of potentially large angle deviations prevalent for light DM masses. 

We find that for sufficiently low energy thresholds, this difference in modelling impacts the DM interaction cross-section sensitivity.  This study evaluates the impact in the context of the QUEST-DMC experiment, which utilises surface-based detectors with superfluid Helium-3 bolometers to search for sub-GeV DM exploiting low energy threshold. At masses below $1~\mathrm{GeV}/c^2$ the deviation between the two frameworks becomes pronounced. The ceiling sensitivity limit for QUEST-DMC on spin-dependent DM-neutron cross-sections is $\sim 3 \times 10^{-24} \mathrm{cm}^2$ using the diffusive framework and approximately doubles with the straight-line path DM scattering. Similarly, for spin-independent DM-nucleon cross-sections, the ceiling limit is $\sim 7.5 \times 10^{-27} \mathrm{cm}^2$ under the diffusive framework and also increases about a factor of two with the straight-line path approximation, within the mass range of $0.025$–$5~\mathrm{GeV}/c^2$.}

\begin{document}
\maketitle
\flushbottom

\section{Introduction}
\label{sec:introduction}

Dark matter (DM) accounts for a significant portion of the universe and has been instrumental in shaping its cosmic evolution. Its indispensable role in the early formation of structures, combined with its gravitational force that prevents galaxies from drifting apart, underscores its importance.

The majority of experimental direct detection searches for DM have traditionally focused on Weakly Interacting Massive Particles with masses well above the proton mass. Despite extensive searches over many years, definitive direct evidence for DM's existence remains elusive. In recent decades, there has been a significant shift in attention towards lighter DM masses, particularly around and below the proton mass. This shift is supported by a range of well-motivated theories~\cite{Zurek:2013wia,Barnes:2020vsc,Hochberg:2014dra,Pospelov:2008jk,Jaeckel:2012mjv,Hall:2009bx}. This includes direct detection searches that primarily focus on identifying elastic DM-nucleus scattering. 

However, detection of DM with masses below a GeV elastically scattering 
necessitates the design of experiments with low-energy thresholds, below the keV scale. Moreover, the recoil energy for sub-GeV DM scattering is inversely proportional to the mass of the target nucleus underscoring the benefits of using low-mass target nuclei. The earlier work~\cite{{QUEST-DMC:2023nug,Autti:2023gxg,Autti:2024awr}} introduced 
the fundamental principles of the Quantum Enhanced Superfluid Technologies for Dark Matter and Cosmology (QUEST-DMC) detector, highlighting its innovative use of superfluid Helium-3 as a target and nanomechanical resonator (NEMS) readout system. This approach takes advantage of the unique properties of the B phase of Helium-3, where thermal excitations, known as quasiparticles, carry the energy deposited by DM interactions. By employing a NEMS made from superconducting metal, QUEST-DMC is able to measure the quasiparticle density, providing a sensitive means of detecting collision events. 

The QUEST-DMC experiment is designed to reach world-leading sensitivity to {\it small} spin-dependent DM-neutron scattering cross-sections, and for spin-independent DM-nucleon scattering, at masses below $0.025~{\rm GeV}/c^2$. This study investigates the sensitivity of QUEST-DMC to scenarios involving {\it large} DM interaction cross-sections, where the experiment approaches its upper sensitivity limit (sensitivity ceiling), and compares its performance to that of competing detection efforts.

\begin{figure}[t]
\centering\includegraphics[width=0.60\linewidth]{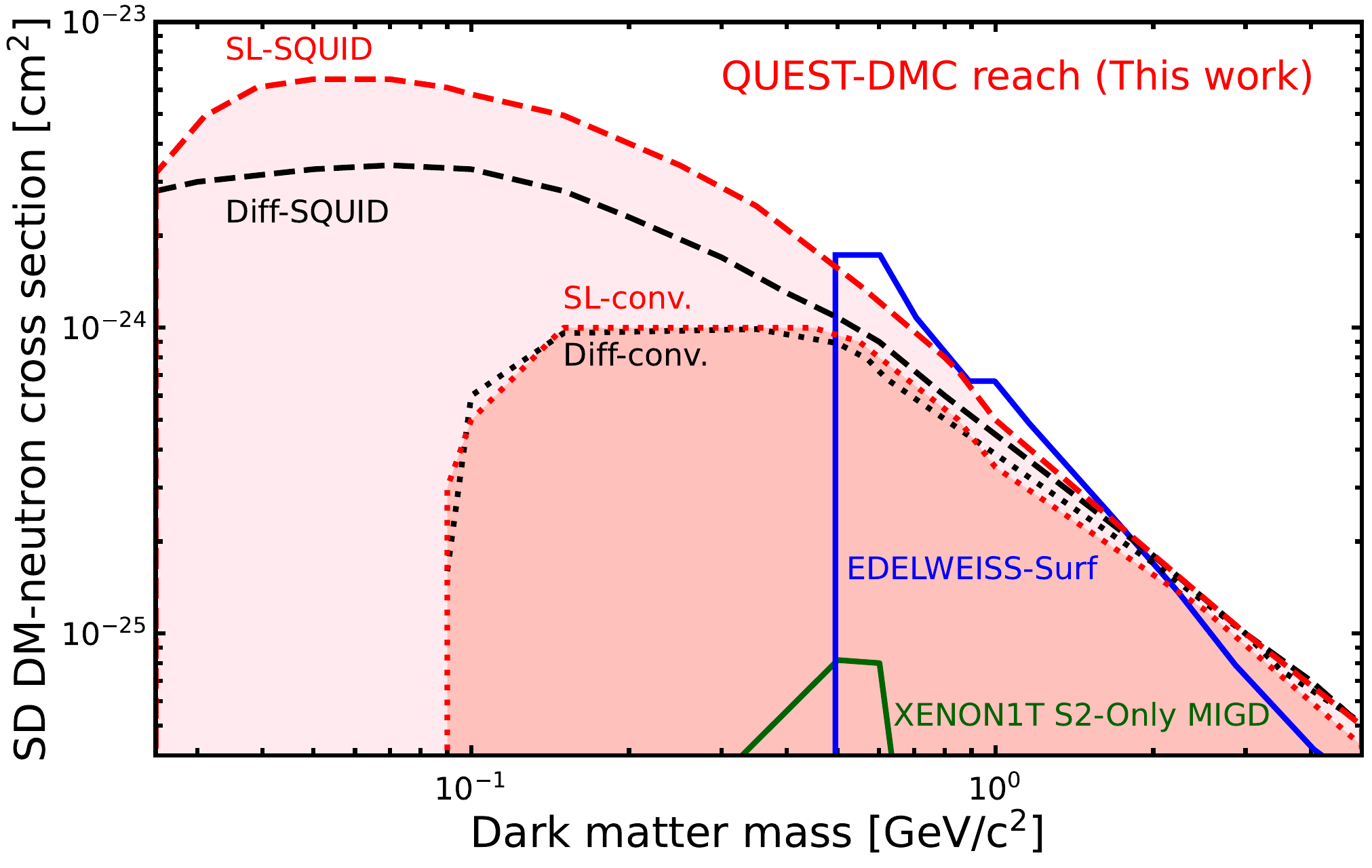}
\caption{Sensitivity predictions of the QUEST-DMC experiment to DM with spin-dependent interactions. The QUEST-DMC sensitivity ceiling is illustrated with two primary trajectories: the straight-line (SL) path, shown in red, and the diffusive (Diff) framework, depicted in black. The dashed lines correspond to SQUID-based readout systems, while the dotted lines denote conventional readout methods.
}
\label{fig:zoomin}
\end{figure}
At large cross-section values the DM flux is attenuated by scattering in the Earth and atmosphere, modifying the velocity distribution at the detector location ~\cite{Collar:1992qc,Collar:1993ss,Hasenbalg:1997hs,Kouvaris:2014lpa,Kouvaris:2015laa,Bernabei:2015nia,Kavanagh:2016pyr}. This effect curtails the observed signal spectra and limits the projected sensitivities of direct detection experiments for large interaction cross-sections-- resulting in a sensitivity ceiling. In Sec.~\ref{sec:NuclearStopping} we calculate the attenuation of the DM flux incident on the QUEST-DMC detector, which is located at the surface. We find that DM particles arrive predominantly from above, as those travelling upwards are effectively blocked due to their scatters in the Earth. We contrast the straight-line approximation used in most studies in the literature with an adaptation of the diffusion framework developed in~\cite{Cappiello:2023hza} to model the attenuation of light DM in the atmosphere. The diffusion model adopted in this work takes into account the potentially large-angle scattering of sub-GeV DM, not modelled in the straight-line approximation. 

Finally, we examine the implications of Earth's atmospheric shadowing on the sensitivities of spin-dependent and spin-independent interactions.  The key result is summarised in Fig.~\ref{fig:zoomin}, showing the expected QUEST sensitivity given QUEST-DMC's two possible readout techniques, described briefly in Sec.~\ref{sec:sensitivity}. We find that QUEST can cover many orders of magnitude of yet unexplored parameter space in the {\it large} cross-section-mass plane, and is in particular competitive for spin-dependent neutron scattering.

\section{Analysis of Direct Detection of Dark Matter} \label{sec:sensitivity} 

The differential event rate for a DM particle of mass $m_{\chi}$ scattering with a target nucleus of mass, $m_{\rm N}$, is
\begin{equation}
\frac{\mathrm{d}R}{\mathrm{d}E_{\rm NR}}=\frac{\rho_{\chi}}{m_{\rm \chi}m_{\rm N}}\int^{\infty}_{v_{\rm min}} \frac{\mathrm{d}\sigma}{\mathrm{d}E_{\rm NR}}\;v\;f(\mathbf{v})\;\mathrm{d}^3\mathbf{v},
\label{eq:Rate}
\end{equation}
where $v_{\rm min}=({m_{\rm N}E_{\rm NR}/2\mu^2_{\rm \chi N}})^{1/2}$ is the minimum DM velocity needed to impart recoil energy of $E_{\rm NR}$ in the detector, $\rho_{\chi}=0.3 \;\rm GeV/c^2$ is the local DM density and $f(\mathbf{v})$ is the local DM velocity distribution in the rest frame of the detector~\cite{Lewin:1995rx,Savage:2006qr} . The Standard Halo Model is assumed, with the DM configured as an isothermal, spherically symmetric halo characterised by an isotropic Maxwell-Boltzmann velocity distribution, which is truncated at the escape velocity, $v_{\rm esc}$, where $\mu_{\rm \chi N}=m_{\chi}m_{\rm N}/(m_{\chi}+m_{\rm N})$ is the reduced mass of the DM-target nucleus system.  

The differential cross-section, $\mathrm{d}\sigma/{\mathrm{d}E_{\rm NR}}$, has in general contributions from both spin-dependent and spin-independent processes of the form
\begin{equation}
\frac{\mathrm{d}\sigma}{\mathrm{d}E_{\rm NR}}=\frac{\mathrm{d}\sigma^{\rm SD}}{\mathrm{d}E_{\rm NR}}+\frac{\mathrm{d}\sigma^{\rm SI}}{\mathrm{d}E_{\rm NR}}.
\end{equation}
In order to compare across experiments and against theory it is assumed that either the spin-dependent or spin-independent process is present (and not both simultaneously).  

Focusing first on the spin-dependent only case for a Helium-3 target, 
the differential cross-section can be written in terms of the spin-dependent DM-neutron scattering cross-section, $\sigma^{\rm SD}_{\rm \chi n}$, as
\begin{equation}
\frac{\mathrm{d}\sigma^{\rm SD}}{\mathrm{d}E_{\rm NR}}=\frac{m_{\rm N}}{2 v^2}\frac{ \sigma^{\rm SD}_{0}}{\mu^2_{\rm \chi N}},
\label{generic-sigSD}
\end{equation}
with
\begin{equation}
\sigma_{0}^{\rm SD} \equiv \frac{\mu^2_{\rm \chi N}}{\mu^2_{\rm \chi n}} \sigma_{\rm \chi n}^{\rm SD}\left[\frac{4}{3} \frac{J_{\rm N} + 1}{J_{\rm N}} \langle{\bf S}\rangle^2 \right],
\label{Eq:Sigma0SD}
\end{equation}
where $\mu_{\rm \chi n}$ is the DM-neutron reduced mass, $m_{\rm N}$ is the nucleus mass (relevant nuclei include Helium-3 at the detector and Nitrogen-14, the dominant spin-dependent atom in the atmosphere), $J_{\rm N}$ is the total nuclear spin ($J_{\rm He3} = 1/2$ and $J_{\rm N14} = 1$), and the mean value of spin for neutrons is $\langle{\bf S_n}\rangle = 1/2$.  

For the spin-independent case, the differential cross-section is given by
\begin{equation}
    \frac{\mathrm{d}\sigma^{\rm SI}}{\mathrm{d}E_{\rm NR}}=\frac{m_{\rm N}\sigma_0^{\rm SI} F^2(E_{\rm NR})}{2v^2\mu_{\rm \chi N}^2},
\end{equation}
where $F^2(E_{\rm NR})$ is a nuclear form factor accounting for the loss of coherence for large momentum transfer. The parameter $\sigma_0^{\rm SI}$ contains the DM interaction details and can be written as~\cite{Kopp:2009qt}
\begin{equation}
    \sigma_0^{\rm SI}=\frac{\left(Zf_{\rm p}+(A-Z)f_{\rm n}\right)^2}{f_{\rm p}^2}\frac{\mu^2_{\rm \chi N}}{\mu^2_{\rm \chi p}}\sigma_{\rm \chi p}^{\rm SI},
\label{Eq:Sigma0SI}
\end{equation}
where $Z$ and $A$ are the atomic and mass numbers, with the most relevant atoms for the spin-independent case being Nitrogen-14 ($Z=7$, $A=14$) and Oxygen-16 ($Z=8$, $A=16$) in the atmosphere, and Helium-3 ($Z=2$, $A=3$) in the detector. The parameters $f_{\rm p,n}$ are the DM-proton/neutron couplings and $\sigma_{\rm \chi p}^{\rm SI}$ is the spin-independent DM-proton cross-section. We assume $f_{\rm p}=f_{\rm n}$ and in the limit of zero momentum transfer, where $F^2(E_{\rm NR})\rightarrow 1$, the differential cross-section simplifies to
\begin{equation}
    \frac{\mathrm{d}\sigma^{\rm SI}}{\mathrm{d}E_{\rm NR}}=\frac{m_{\rm N} A^2}{2v^2\mu_{\rm \chi p}^2}\sigma_{\rm \chi p}^{\rm SI}.
\end{equation}

Substituting these for Helium-3 into the differential event rate for spin-dependent and spin-independent scattering gives
\begin{eqnarray}
\label{drdeSD}
    \frac{\mathrm{d}R^{\rm SD}}{\mathrm{d}E_{\rm NR}}=
    \frac{\rho_{\chi}\,\sigma^{\rm SD}_{\rm \chi n}}{2 m_{\chi}\,\mu^2_{\rm \chi n}}
    \int^{\infty}_{v_{\rm min}} \frac{1}{v}\;f(\mathbf{v})\;\mathrm{d}^3\mathbf{v}, \\
\label{drdeSI}
    \frac{\mathrm{d}R^{\rm SI}}{\mathrm{d}E_{\rm NR}}=
    \frac{9\rho_{\chi}\sigma^{\rm SI}_{\rm \chi p}}{2 m_{\chi}\,\mu^2_{\rm \chi p}}
    \int^{\infty}_{v_{\rm min}} 
    \frac{1}{v}\;f(\mathbf{v})\;\mathrm{d}^3\mathbf{v}.
\end{eqnarray}

Detector response effects, including energy partition fluctuations, shot noise, and readout noise, are incorporated into the predicted differential event rates as a function of energy, following the model described in~\cite{QUEST-DMC:2023nug}.  
The lower limit of the velocity integral is determined by the recoil energy threshold of the QUEST-DMC detector. 
To reach low energy threshold, QUEST-DMC uses two different readout techniques: one is based on a conventional cold transformer readout, which is a sensitive amplifier designed to operate at cryogenic temperature. It amplifies the small electrical signal generated during particle interactions within the detector and achieves an energy threshold of ${\rm 31~eV}$ at 95\% confidence level. The other method uses Superconducting Quantum Interference Device (SQUID), which is an extremely sensitive magnetic flux quantum sensor operating at superconducting temperature. This method is capable of detecting much smaller signals and achieves a significantly lower energy threshold of ${\rm 0.51~eV}$ at 95\% confidence level.

\section{Attenuation of the Dark Matter Flux}
\label{sec:NuclearStopping}
The attenuation effect arises from the scattering of DM particles as they traverse the Earth and its atmosphere. This phenomenon becomes increasingly significant for DM particles with large scattering cross-sections, where frequent interactions with nuclei can result in energy loss and ultimately complete attenuation such that no DM flux reaches the detector. 
As the elastic scattering cross-section increases, there is a point at which any DM particles travelling through the Earth are stopped, while the DM flux traversing only the atmosphere can still reach the detector. Thus for the largest cross-section we assume the DM flux is predominantly incident from above.

To model these effects two methods are examined: the straight-line path approximation and a diffusion framework.

The straight-line path approximation assumes that DM particles follow a straight-line trajectory from their point of origin, deviating by a negligible amount in collisions with nuclei. This approach simplifies calculations by treating interactions as isolated events along an unaltered path. While effective for heavy DM, where we expect minimal deviations in path, this model can overestimate the number of particles reaching the detector when applied to light DM as it neglects the cumulative effects of multiple scatterings with directional changes and thus larger energy redistribution.

The diffusion framework, by contrast, models DM particle interactions as a series of isotropic scatterings, simulating a random walk through the atmosphere. This approach accounts for both angular and energy deviations resulting from successive interactions, providing a more realistic representation of DM trajectories. It is particularly relevant for sub-GeV DM, where frequent collisions and energy loss dominate, leading to significant attenuation and redistribution of the velocity spectrum. 

In this work, these methodologies are applied to refine predictions of detection rates and sensitivity limits for QUEST-DMC. We show that while the straight-line path approximation offers a computationally efficient baseline for heavy DM scenarios, the diffusion framework enhances precision for light DM, where the opacity of the Earth's atmosphere becomes a critical factor. By comparing the two methods, we refine the modelling of the interplay between the scattering cross-section, DM mass, and detector response, ensuring robust interpretation of experimental results and limit setting. This section describes the implementation and implications of these approaches for the QUEST-DMC experiment.

\subsection{Straight-Line-Approximation}
\label{sec:VelDist}

In this section, we examine the integration over the velocity distribution of DM, $f(\mathbf{v})$. This integration includes contributions from all particles that possess a sufficient DM speed to induce a recoil energy $E_{\rm NR}$ above the detector energy threshold in QUEST-DMC. 

The velocity distribution in Eq. \eqref{eq:Rate} is given by
\begin{align}
\label{eq:veldist}
f(\mathbf{v}) = \frac{1}{N_0} \exp\left(-\frac{\left|\mathbf{v} - \langle \mathbf{v}_\chi \rangle\right|^2}{v_0^2}\right) \Theta(v_\mathrm{esc} - \left|\mathbf{v} - \langle\mathbf{v}_\chi\rangle\right|)\,,
\end{align}
representing the unattenuated velocity distribution of DM particles arriving at the Earth with the normalisation constant $N_0$ given by
\begin{align}
\label{norm-bd}
N_0=\left( \sqrt{\pi} v_0\right)^3 \;\left[ \mathrm{erf}\left(\frac{v_\mathrm{esc}}{v_0}\right) \;-\;
\frac{2}{\sqrt{\pi}}\frac{v_\mathrm{esc}}{v_0} \exp\left(-\frac{v_\mathrm{esc}^2}{v_0^2}\right)\right],
\end{align}
with $v_0 = 220.617 \,\mathrm{km/s}$ and an escape speed
in the Galactic frame $v_{\rm esc}= 544 ~\rm km/s$~\cite{DMStat_2021}.

\begin{figure}[t!]
\centering
\includegraphics[width=0.35\textwidth,]{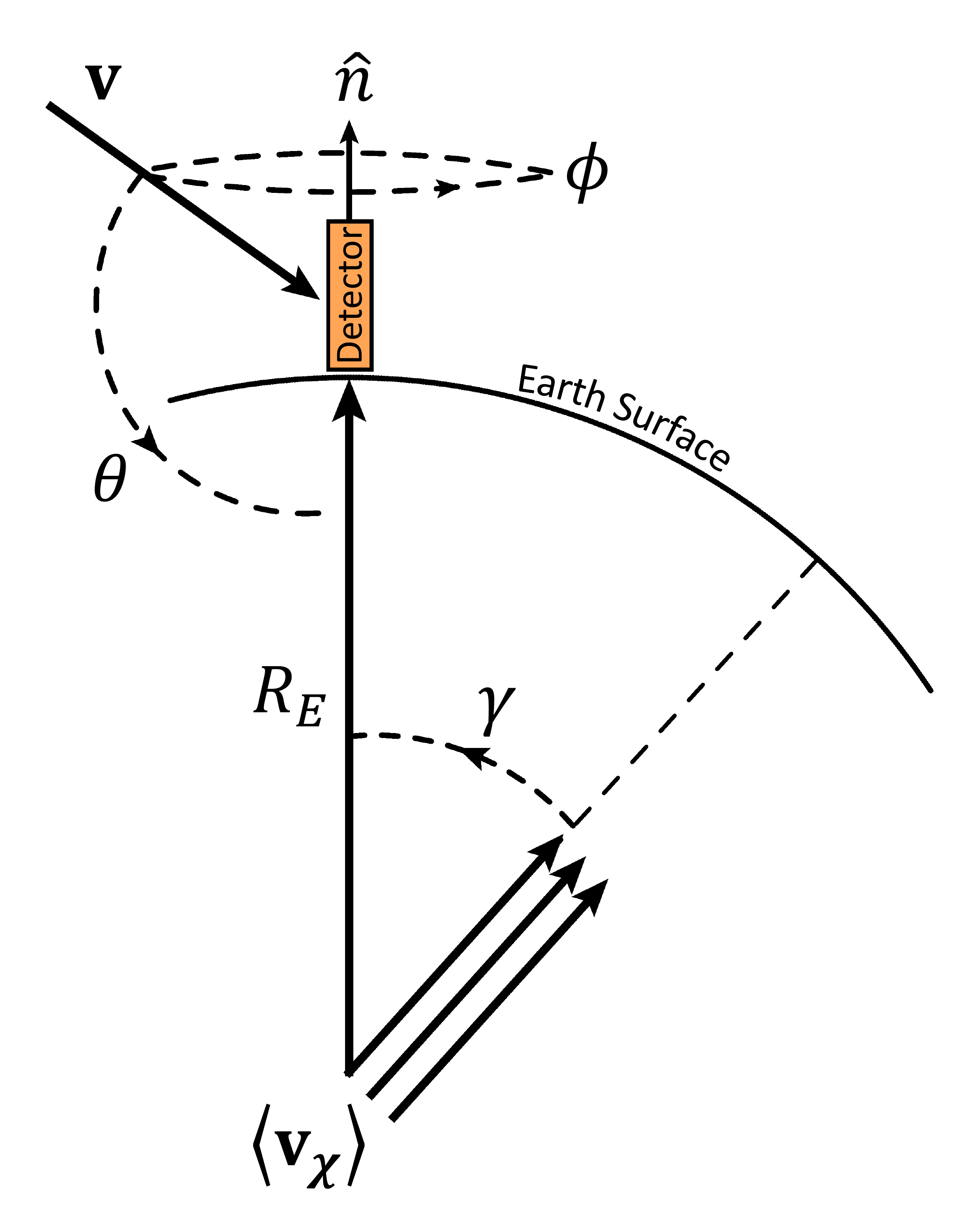}
\caption{Coordinate system for DM scattering from the top of the atmosphere towards a detector located on the Earth's surface. The incoming DM particle has a velocity vector $ \mathbf{v}$, which is characterised by the polar angles $\theta$ and $\phi$. The average DM flux velocity vector, denoted by $\langle\mathbf{v}_\chi \rangle$, forms an angle $\gamma$ with respect to the detector's normal vector $\hat{n}$. The Earth's radius is represented by $R_E$.}
\label{fig:geo}
\end{figure}

This distribution can be altered if the DM undergoes significant interactions in the atmosphere or in the Earth itself. The modification depends on where the detector is situated, for example, whether it is on the surface or underground and the path taken by the DM particles to reach the detector. We assume that the unattenuated velocity distribution $f(\mathbf{v}_{\rm i})$ (at the upper atmosphere, before scattering) follows the Maxwell-Boltzmann form given in Eq.~\eqref{eq:veldist}. We employ spherical coordinates to describe the incoming DM velocity $\mathbf{v}$, with the usual polar angles $\theta$ and $\phi$ as shown in Fig.~\ref{fig:geo} (see Ref.~\cite{Kavanagh:2016pyr}). In these coordinates, $\mathbf{v}$ can be written as  
$$
\mathbf{v} = v \,\bigl(\hat{x}\,\sin\theta\,\cos\phi + \hat{y}\,\sin\theta\,\sin\phi + \hat{z}\,\cos\theta\bigr).
$$
The average DM velocity (which reflects the motion of the Earth through the DM halo) lies in the $x\!-\!z$ plane, and is defined as
$$
\langle \mathbf{v}_\chi \rangle = v_E \,\bigl(\hat{x}\,\sin\gamma + \hat{z}\,\cos\gamma\bigr),
$$
where $v_E = 232 \,\mathrm{km/s}$ is the magnitude of the Earth's velocity relative to the halo, and $\gamma$ is the angle between $\langle \hat{\mathbf{v}}_\chi \rangle$ and the position of detector $\hat{n}$.
To characterise the geometry, we assume $\delta$ as the angle between the vectors $\mathbf{v}$ and $\langle \mathbf{v}_\chi \rangle$. Its cosine becomes
$$
\cos\delta = \sin\theta\,\cos\phi\,\sin\gamma + \cos\theta\,\cos\gamma.
$$
The dependence of quantities like $|\mathbf{v} - \langle \mathbf{v}_\chi \rangle|$ on $\theta$, $\phi$, and $\gamma$ is fully encapsulated by $\delta$. However, the final DM velocity at the detector remains independent of the azimuthal angle $\phi$.

The evaluation of the final speed distribution $f(v_{\rm f})$ at the detector for spin-independent and spin-dependent interactions is~\cite{Kavanagh:2017cru} 
\begin{align}
\label{eq:modifyF}
f(v_{\rm f}^{\rm int},\gamma)  &\,=\,  \oint \, v_{\rm i}^2 \, f(\mathbf{v}_{\rm i},\gamma) \frac{\mathrm{d} v_{\rm i}}{\mathrm{d} v_{\rm f}^{\rm int}} \, \mathrm{d}^2\hat{\mathbf{v}}
\nonumber \\
& \,=\, \int_{-1}^{1} \int_{0}^{2\pi}\, v_{\rm i}^2 \, f({v}_{\rm i}, \gamma, \cos\theta, \phi) \frac{\mathrm{d} v_{\rm i}}{\mathrm{d} v_{\rm f}^{\rm int}} \mathrm{d}\phi\,\mathrm{d}(\cos\theta)\,
\nonumber \\
& \,=\, \int_{-1}^{1} \, v_{\rm i}^2 \, f({v}_{\rm i}, \gamma, \cos\theta) \frac{\mathrm{d} v_{\rm i}}{\mathrm{d} v_{\rm f}^{\rm int}} \,\mathrm{d}(\cos\theta)\,,
\end{align} where int $\equiv$ SD or int $\equiv$ SI, and in the above relation the two angular integrations can be performed separately with $f(v_{\rm i},\,  \gamma \, ,\cos\theta) = \int_{0}^{2\pi} f(v_{\rm i},\, \gamma \, , \cos\theta,\,\phi)\,\mathrm{d}\phi$. 

The velocity distribution of DM particles is influenced by their path length through the atmosphere. This relationship can be described using angular coordinates that define both the DM trajectory and the detector's orientation relative to the primary direction of the DM flux.

For a DM particle arriving at the detector at an angle $\theta$ with respect to the vertical direction (see Fig.~\ref{fig:geo}), the straight-line path length $L$ through the atmosphere is given by
\begin{equation}
L \,=\, R_E \cos\theta + d
\end{equation}
with
\begin{equation}
d \,=\, \Big[(R_E + H)^2 - (R_E \sin\theta)^2\Big]^{1/2}\,,
\end{equation}
where $ R_E \approx 6371\,\mathrm{km} $ is the Earth radius and $ H \approx 80 \,\mathrm{km} $ is the vertical extent of the atmosphere.

To accurately compute the scattering rate for DM particles, we must consider the radial dependence of the nuclear density in the atmosphere. 
The number density with the assumption of an exponential atmospheric model is
\begin{equation}
n_{\rm N}(\mathbf{r}) \,= \, n_0~ {\rm exp}\Big(-{(\mathbf{r} - R_E) \over H}\Big).
\end{equation}
where $n_0$ is the sea-level number density. For a DM particle that has travelled a distance $d_{\rm SLP}$ along its trajectory, measured from the top of the atmosphere, the radial distance from the centre of Earth is:
\begin{equation}
r = \Big[(R_E + H)^2 + (d_{\rm SLP}-d)^2 - d^2\Big]^{1/2}\,.
\end{equation}
Thus, the difference between the final speed $v_{\rm f}$ (interaction dependent) from the initial speed $v_{\rm i}$ (at the top of the atmosphere) along a given trajectory is:
\begin{equation}
\label{eq:vi-vf}
\Delta v^{\rm int} = \int_0^{L} \frac{\mathrm{d}v^{\rm int}}{\mathrm{d}d_{\rm SLP}}(v, r) \,\mathrm{d}d_{\rm SLP}\,,
\end{equation}
which depends on both $r$ and $v$. 

To determine the final velocity of DM particles incident on the detector, we adopt the ``nuclear stopping'' methodology first introduced in Refs.~\cite{Starkman:1990nj,Kavanagh:2017cru,Davis:2017noy}.  

In Sec.~\ref{sec:sensitivity}, we illustrated the differential cross-section for general contributions from spin-dependent and spin-independent interactions with Helium-3. However, as DM particles traverse the atmosphere, containing various nuclear species denoted by index $\mathrm{N}$, they may interact and lose energy. 
Since the detector is located on the surface, particles start from the top of the atmosphere and end at the Earth's surface.  
The usual justification for assuming a straight-line trajectory between these two points is that the average distance particles travel between collisions is significantly larger than the thickness of media like Earth's air layers. This suggests that collisions are relatively infrequent. 
Furthermore, these particles are assumed to be moving at high velocities, meaning any collisions would likely result in only minor directional changes. 
In the event of an elastic collision, where energy loss is minimal, particles largely maintain their initial directions. Thus one can postulate that their overall scattering path is continuous and predominantly straight-line. Under this assumption, the rate of change of the average DM energy, which in general has contributions from both spin-dependent and spin-independent processes, is of the form
\begin{equation}
 \frac{\mathrm{d}\langle E_\chi \rangle}{\mathrm{d}t}=\frac{\mathrm{d}\langle E_\chi \rangle^{\rm SD}}{\mathrm{d}t}+\frac{\mathrm{d}\langle E_\chi \rangle^{\rm SI}}{\mathrm{d}t}.
\end{equation}

Now considering $n_{\rm N}(\mathbf{r})$ as the number density of nuclei $\mathrm{N}$ at position $\mathbf{r}$ (based on the International Standard Atmosphere~\cite{ISA}, which encompasses up to an altitude of 80 km) the rate for each contribution (${\rm int}\equiv$ SD or ${\rm int} \equiv$ SI) is
\begin{eqnarray}
   \frac{\mathrm{d}\langle E_\chi \rangle^{\rm int}}{\mathrm{d}t} = 
   -2v \sum_{\rm N} n_{\rm N}(\mathbf{r}) 
   \frac{\mathrm{d}\sigma^{\rm int}}{\mathrm{d}E_{\rm NR}} 
   \left( \frac{\mu_{\rm \chi N}^2 v^2}{m_{\rm N}} \right)^2\,.
   \label{eq:dExdt}
\end{eqnarray}
Taking the DM kinematic energy as $E_\chi = m_\chi v^2/2$, we can obtain the trajectory of DM  as
\begin{align}
\label{Eq:dvdDint}
 \frac{\mathrm{d}v^{\rm int}}{\mathrm{d} d_{\rm SLP}} &= -\frac{v}{m_\chi} \sum_{\rm N} n_{\rm N} (\mathbf{r}) \frac{\mu_{\rm \chi N}^2}{m_{\rm N}} \sigma_{0}^{\rm int}\,.
\end{align}
where $\sigma_{0}^{\rm SI}$ and $\sigma_{0}^{\rm SD}$ are defined in Eqs.~\eqref{Eq:Sigma0SD} and \eqref{Eq:Sigma0SI}, respectively. We note that for spin-dependent DM interactions, the summation considers only Nitrogen-14 as the contribution from Oxygen-17 in the atmosphere is negligible. 
By substituting Eq.~\eqref{Eq:dvdDint} into Eq.~\eqref{eq:vi-vf} the deviation $\Delta v^{\mathrm{int}}$ can be parametrised.
\begin{figure}[t]
\centering
\includegraphics[width=0.5\linewidth]{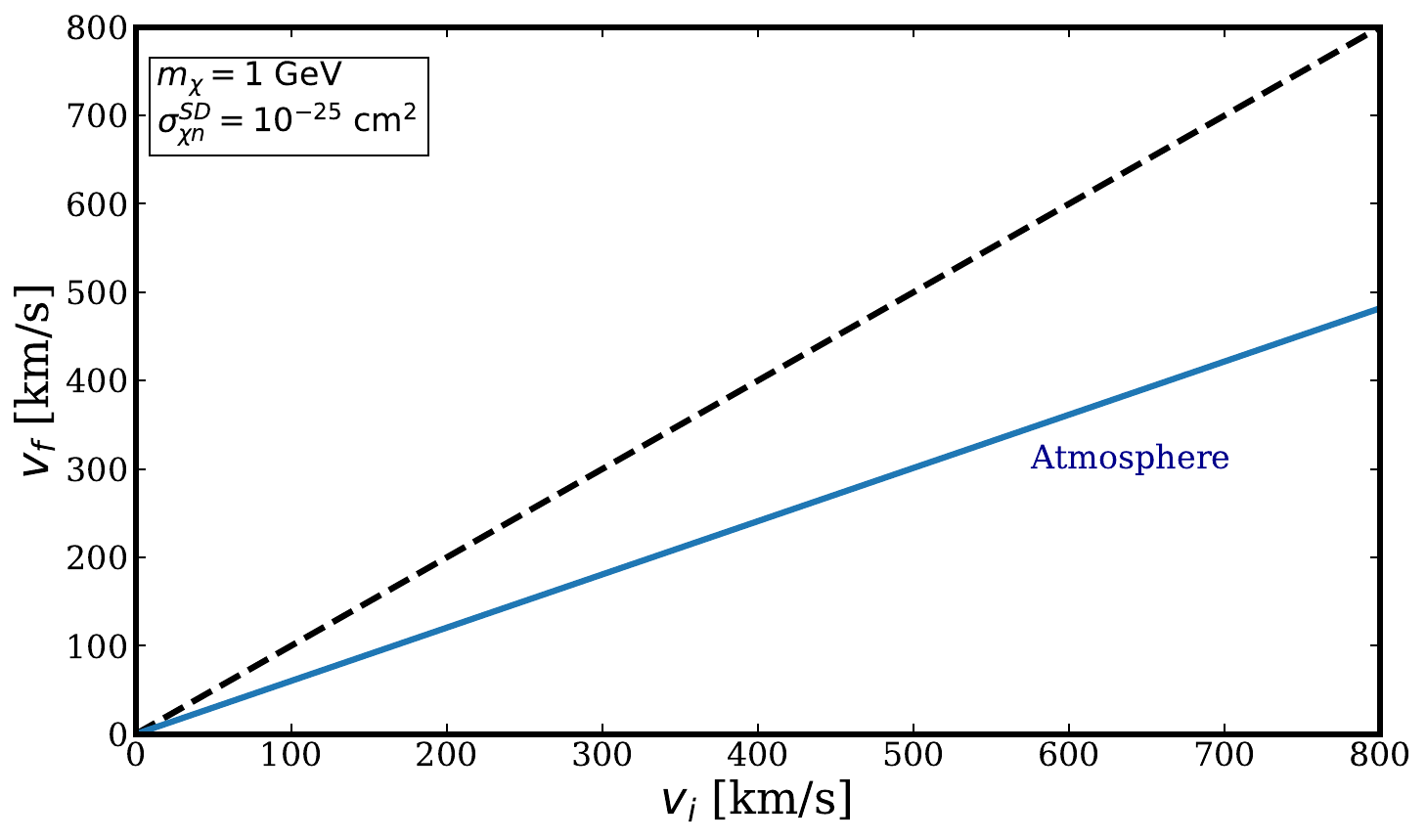}
\caption{Example of mapping the initial speed of DM particles to the final speed at the detector for spin-dependent interactions. The dashed, black line denotes the DM speed at the top of the atmosphere and the solid, blue line denotes the final DM speed after propagation through the atmosphere.}
\label{fig:vivf}
\end{figure}
Fig.~\ref{fig:vivf} presents a graphical representation of the changes in speed as DM propagates through the atmosphere, illustrating how initial DM velocities are mapped to their final velocities upon reaching the detector at the Earth's surface.

These effects significantly modify the velocity distribution for both spin-dependent and spin-independent interactions. Fig.~\ref{fig:Fv} elucidates the distribution of DM velocities as a function of their final speeds upon reaching the detector for both cases. 

\begin{figure*}[t]
\centering
\includegraphics[width=0.5\textwidth]{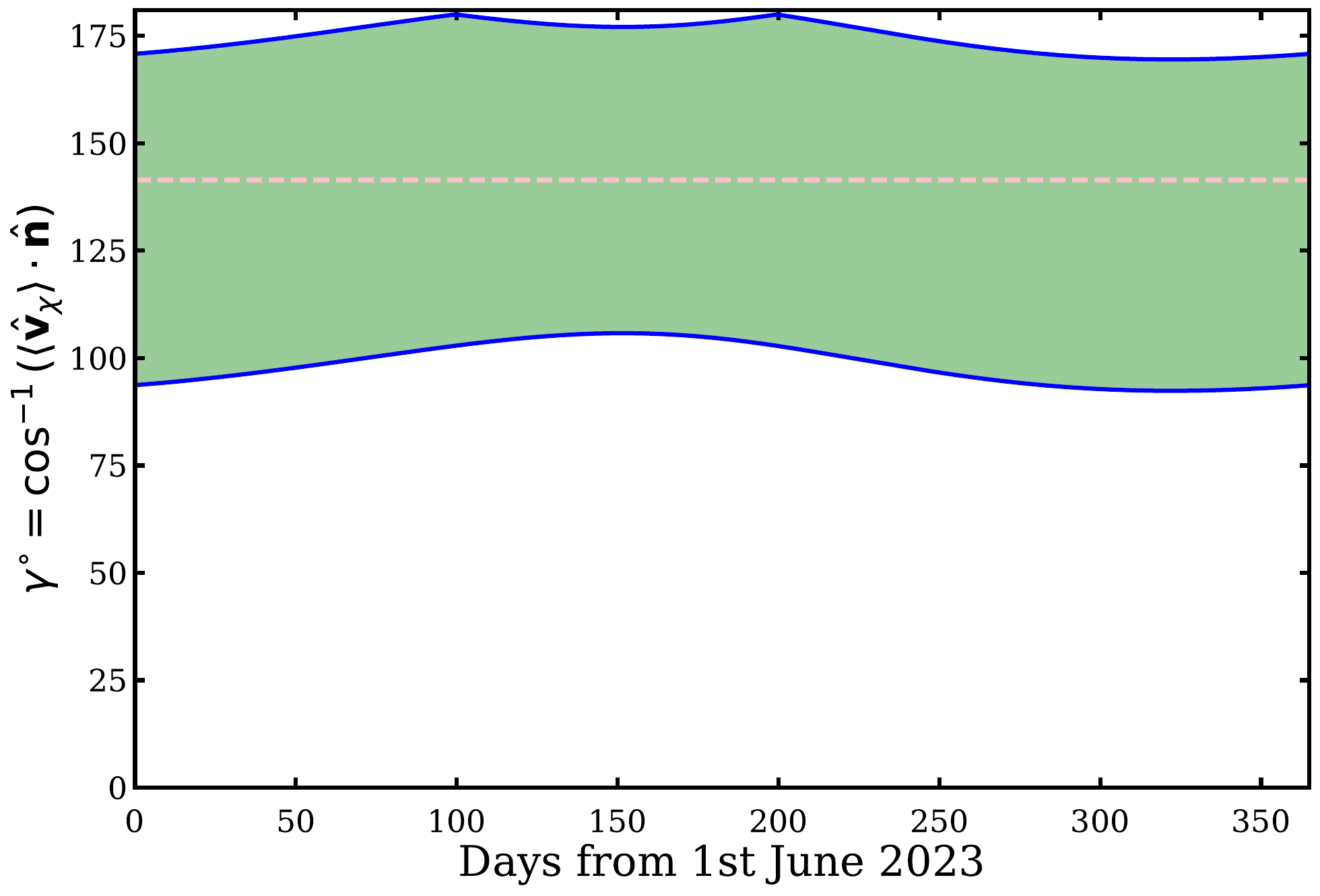}
\caption{The green band shows the daily range of the values of the angle $\gamma$ for one year of QUEST-DMC operation, for the detector located at Royal Holloway, University of London. The dashed yellow line shows the average value of $\gamma$ in this time period. This average value is used as input in the sensitivity calculations in this work.}
\label{Gamma_RHUL_2024}
\end{figure*}

The value of $\gamma$ used in calculating  Fig.~\ref{fig:Fv} is displayed in Fig.~\ref{Gamma_RHUL_2024}, for a 1-year period. The shaded green band shows the maximum and minimum values of 
$\gamma=\arccos\bigl(-\cos\theta_l \,\cos \omega t\,\sin \alpha - \sin\theta_l \,\cos\alpha\bigr)$
for the QUEST-DMC location (latitude $51.43^\circ$ N, longitude $-0.56^\circ$ W) over the course of a year. Here, $\omega = 2\pi/\mathrm{day}$ denotes the angular velocity of the Earth's rotation, and $\alpha$ is the angle between the Earth's velocity vector, $\vec{v}_E$, and the angular velocity vector, $\vec{\omega}$, in the Galactic Rest Frame (varies between $36.3^\circ$ and $49.3^\circ$ throughout the year \cite{Kouvaris:2014lpa}). 
The dashed yellow line in Fig.~\ref{Gamma_RHUL_2024} shows the average value of $\gamma$ used in our calculations throughout this work. 

\subsection{Diffusion Framework}
In this section we describe an alternative method to obtain the attenuation of the DM flux incident on the detector at the surface. While the straight-line approximation is well established for heavy DM, it is physically not justified for sub-GeV DM as it is significantly lighter than the Earth/atmospheric nuclei it scatters with on its path to the detector, and thus this approximation may result in inaccurate results for the DM ceiling, as discussed in Ref.~\cite{Cappiello:2023hza}. 
Instead, DM may scatter isotropically in the lab frame upon colliding with Earth/atmospheric nuclei. In this work we extend the framework developed in~\cite{Cappiello:2023hza} for modelling attenuation of light DM in the atmosphere. To this end we compute probabilities of individual DM particles reaching the detector analytically, incorporating a range of directions of incoming DM, considering up to $n$ scatterings. This calculation extends upon Ref.~\cite{Cappiello:2023hza} by explicitly accounting for the angular distribution of the incoming DM flux.

\begin{figure*}[t]
\includegraphics[width=0.5\textwidth]{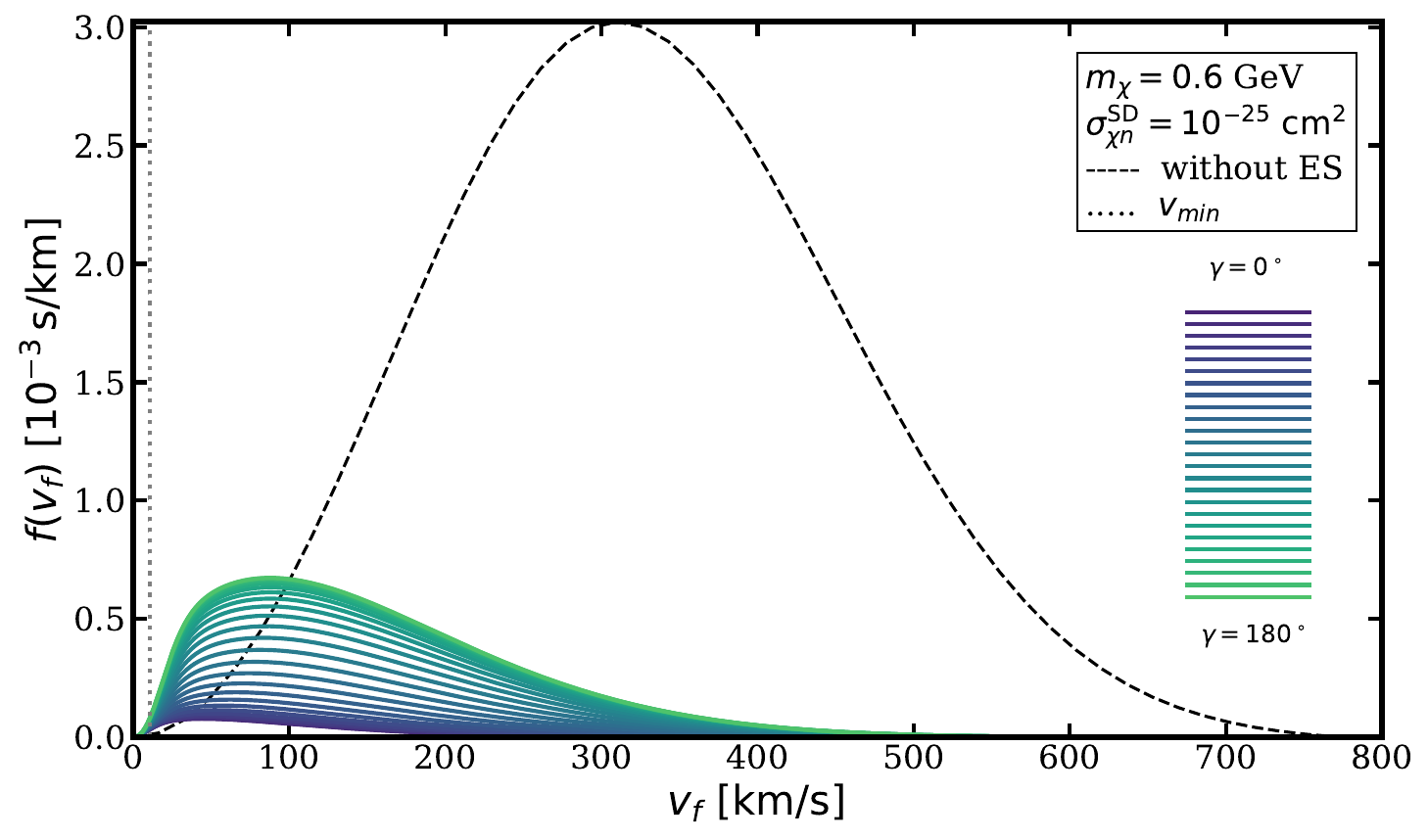}
\includegraphics[width=0.5\textwidth]{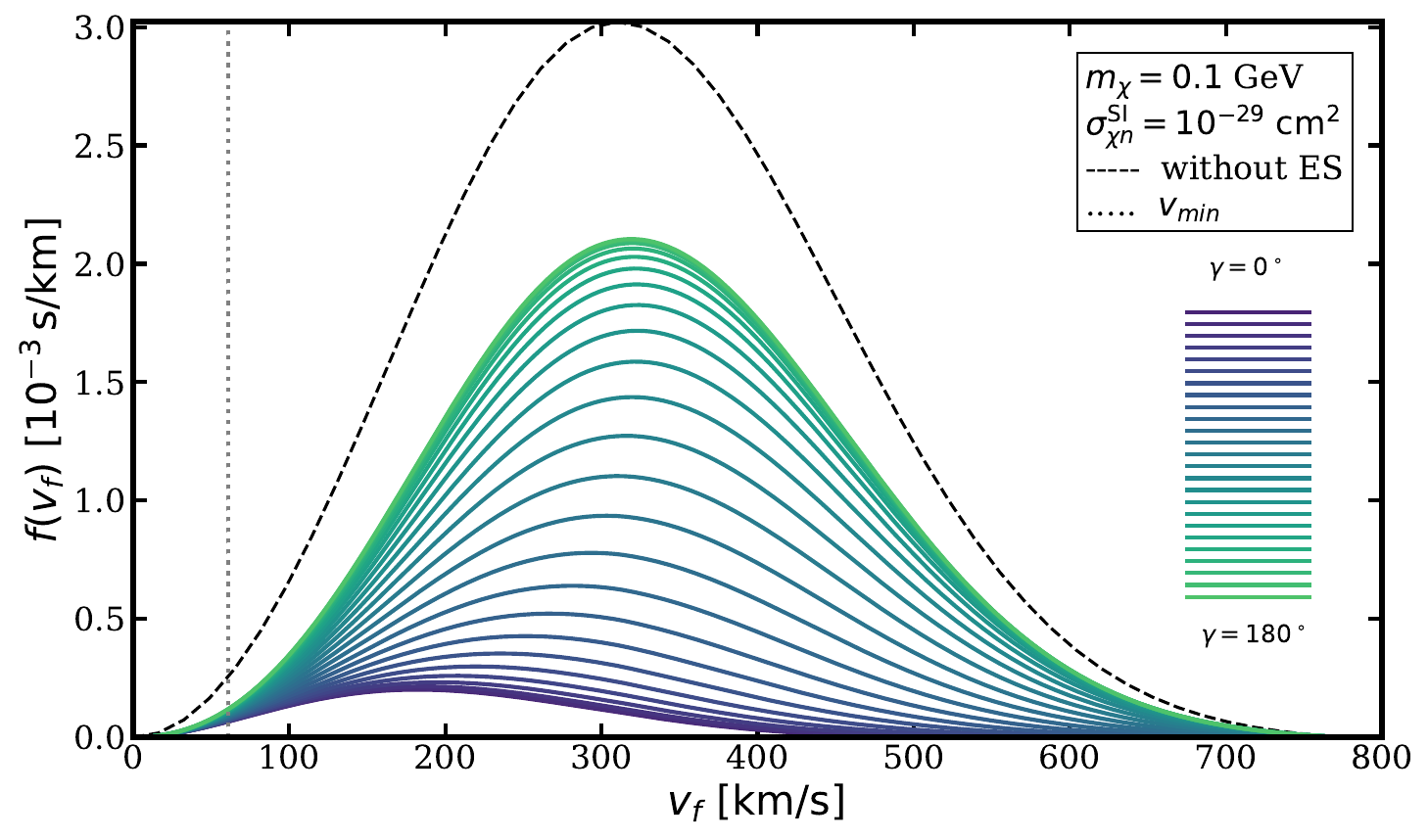}
\includegraphics[width=0.5\textwidth]{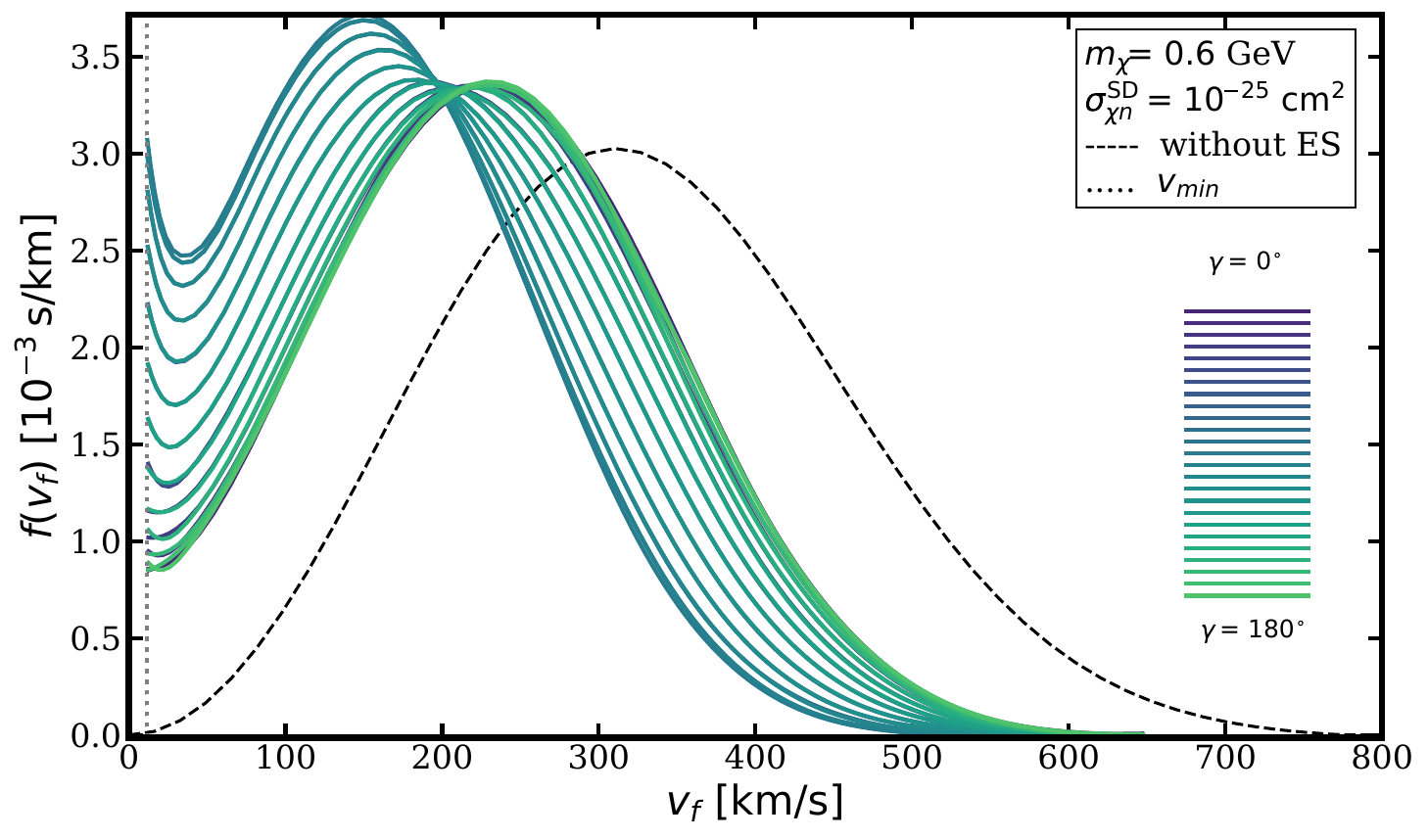}
\includegraphics[width=0.5\textwidth]{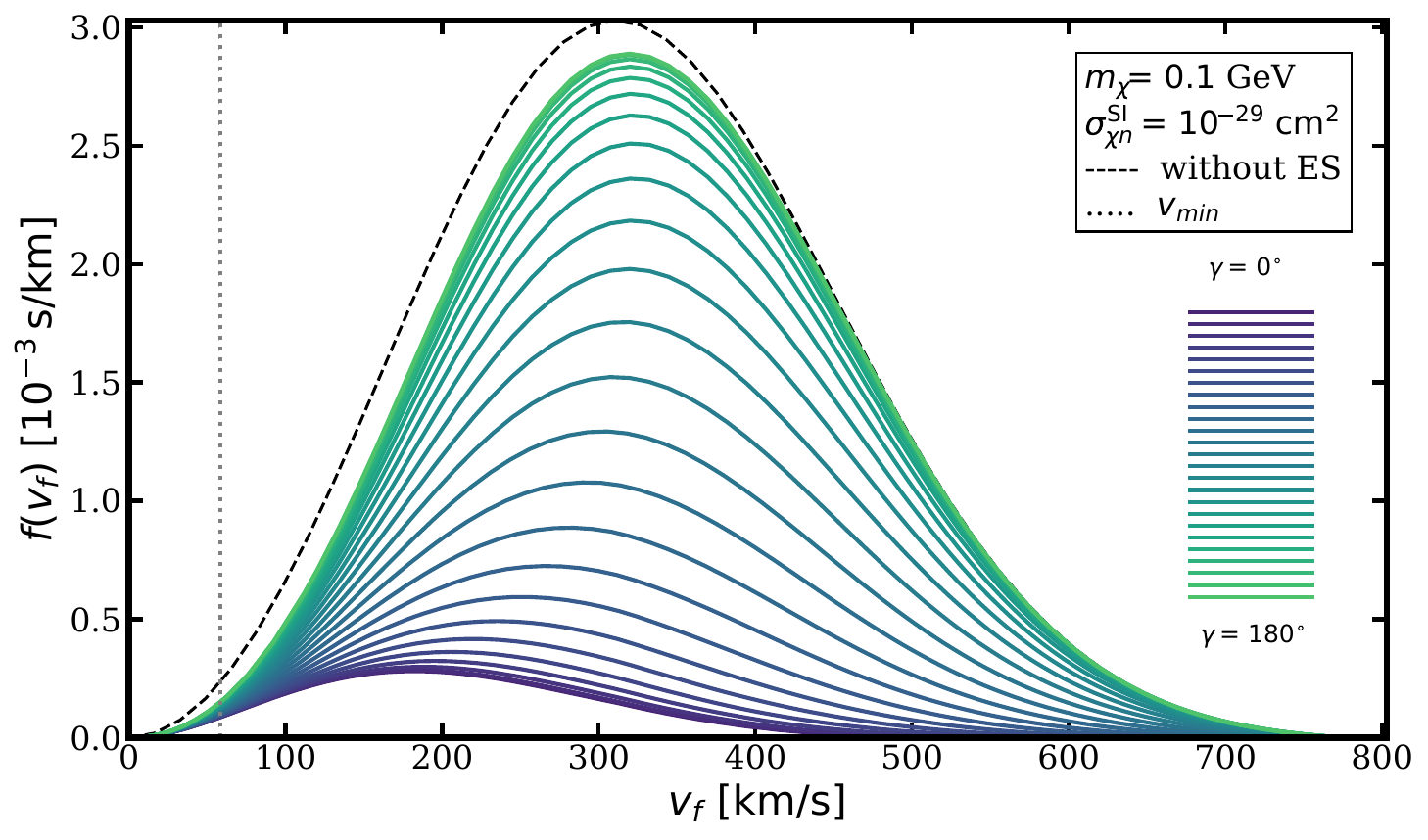}
\caption{Examples of velocity distribution as a function of final velocity for spin-dependent (left) and spin-independent (right) interactions. The top-row plots show the velocity distributions obtained in the extended analytic diffusive model, while the bottom-row plots show the straight-line-path approximation. The dashed line denotes the DM speed at the top of the atmosphere and coloured lines indicate velocity distribution associated with different values of $\gamma$ for spin-dependent with $m_\chi=0.6$ GeV, $\sigma^{\rm SD}_{\rm \chi n}=10^{-25}$cm$^2$ and $m_\chi=0.1$ GeV, $\sigma^{\rm SI}_{\rm \chi n}=10^{-29}$cm$^2$. The grey vertical line indicates the minimal velocity $v_{\text{min}}$, which corresponds to the lower limit on the velocity required to produce a nuclear recoil above the detection threshold in QUEST-DMC with the SQUID readout. 
}
\label{fig:Fv}
\end{figure*}

At the DM mass and cross-section values we are interested in, the Earth is completely opaque to DM particles. To take this effect into account, we integrate the incoming DM flux over the upper half-sphere above the detector. We note that the fraction of the total DM flux in the upper half-sphere varies with the season, as the average DM flux impacts our planet at different angles over the year; and, that at the large cross-section values considered the effects of Earth curvature are negligible. We thus obtain the initial velocity distribution for our treatment from the following integral:
\begin{align}
\label{eq:modifyF}
f_{1/2}(v_{\rm i},\gamma)  &\,=\,  \int_{\text{half-sphere}} \, v_{\rm i}^2 \, f(\mathbf{v}_{\rm i},\gamma) \, \mathrm{d}^2\hat{\mathbf{v}}
\nonumber \\
& \,= \int_{-1}^{0} \, v_{\rm i}^2 \, f(v_{\rm i}, \gamma, \cos\theta)  \,\mathrm{d}(\cos\theta)\,.
\end{align}
\begin{figure}
\centering    \includegraphics[width=0.5\linewidth]{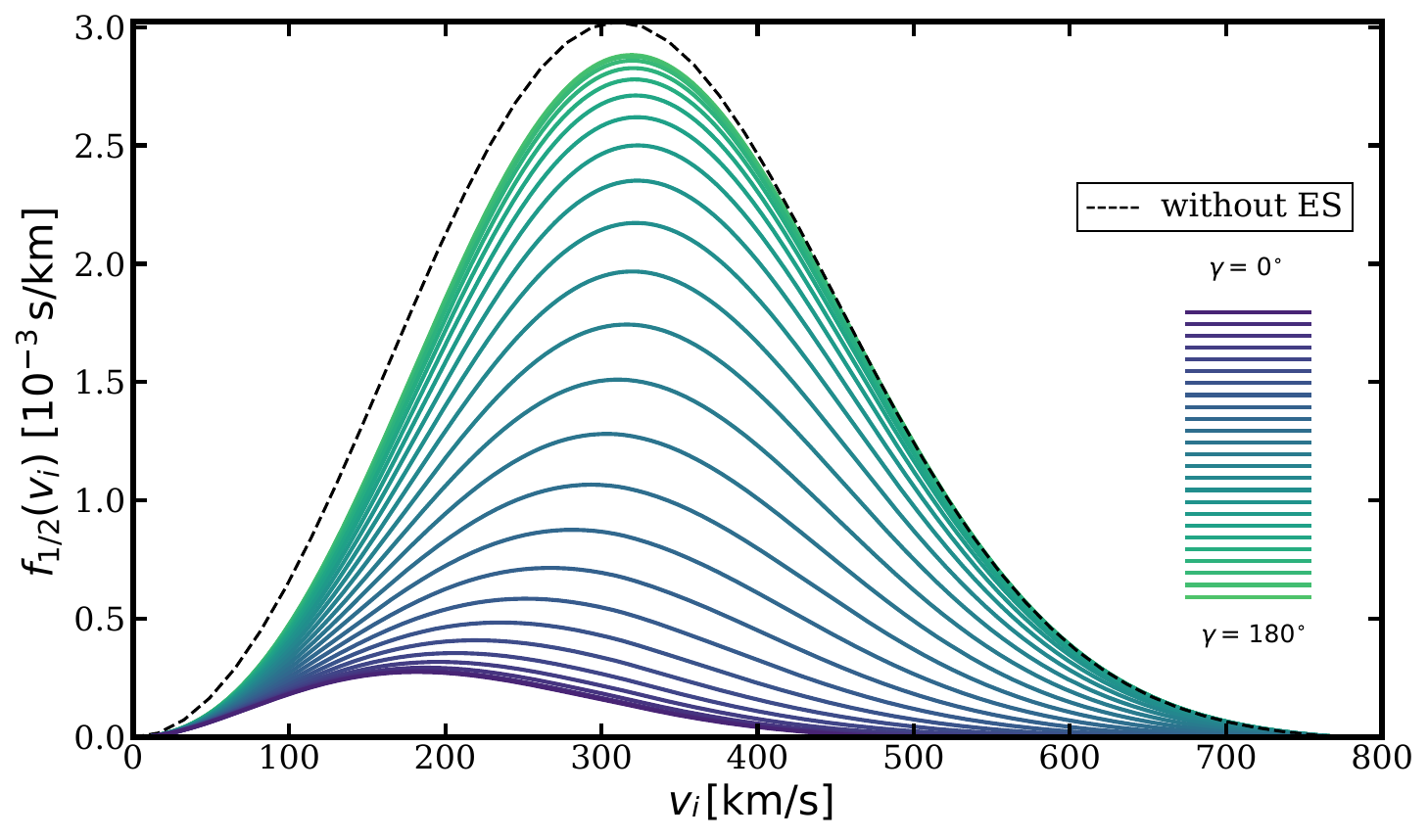}
\caption{The initial DM velocity distribution as a function of the seasonal parameter $\gamma$ under the assumption of the Earth being fully opaque to DM. }
    \label{fig:shadow}
\end{figure}
In Fig.~\ref{fig:shadow}, we show the
DM distribution $f_{1/2}(v_{\rm i},\gamma)$ corrected by the Earth shadow as a function of the seasonal parameter $\gamma$. This initial DM distribution is then passed through the diffusion framework to obtain the final DM velocity distribution at the detector, shown in Fig.~\ref{fig:Fv}. We now summarise the analytic diffusion algorithm.
 
Initially, a fraction of DM particles reach the detector without scattering, for a given mean free path $\lambda$ and for a distance travelled $z$. The probability density for these in terms of the upper incomplete gamma function $\Gamma(0,x)$ is given by~\cite{Cappiello:2023hza}
\begin{equation}
    P_{\text{initial}}(z) = \frac{1}{\lambda}e^{-z/\lambda} = \frac{1}{\lambda}\Gamma(0,z/\lambda) \,,
\end{equation}
assuming that DM arrives isotropically (from above $\gamma=180^\circ$) and neglecting backscattering. This implies that without attenuation, we rely on the Boltzmann distribution (Eq. \eqref{eq:modifyF}) where we are incorporating the angular dependence on $\gamma$ to account for the direction of the incoming DM. 

\begin{figure*}[t]
\includegraphics[width=0.5\textwidth]{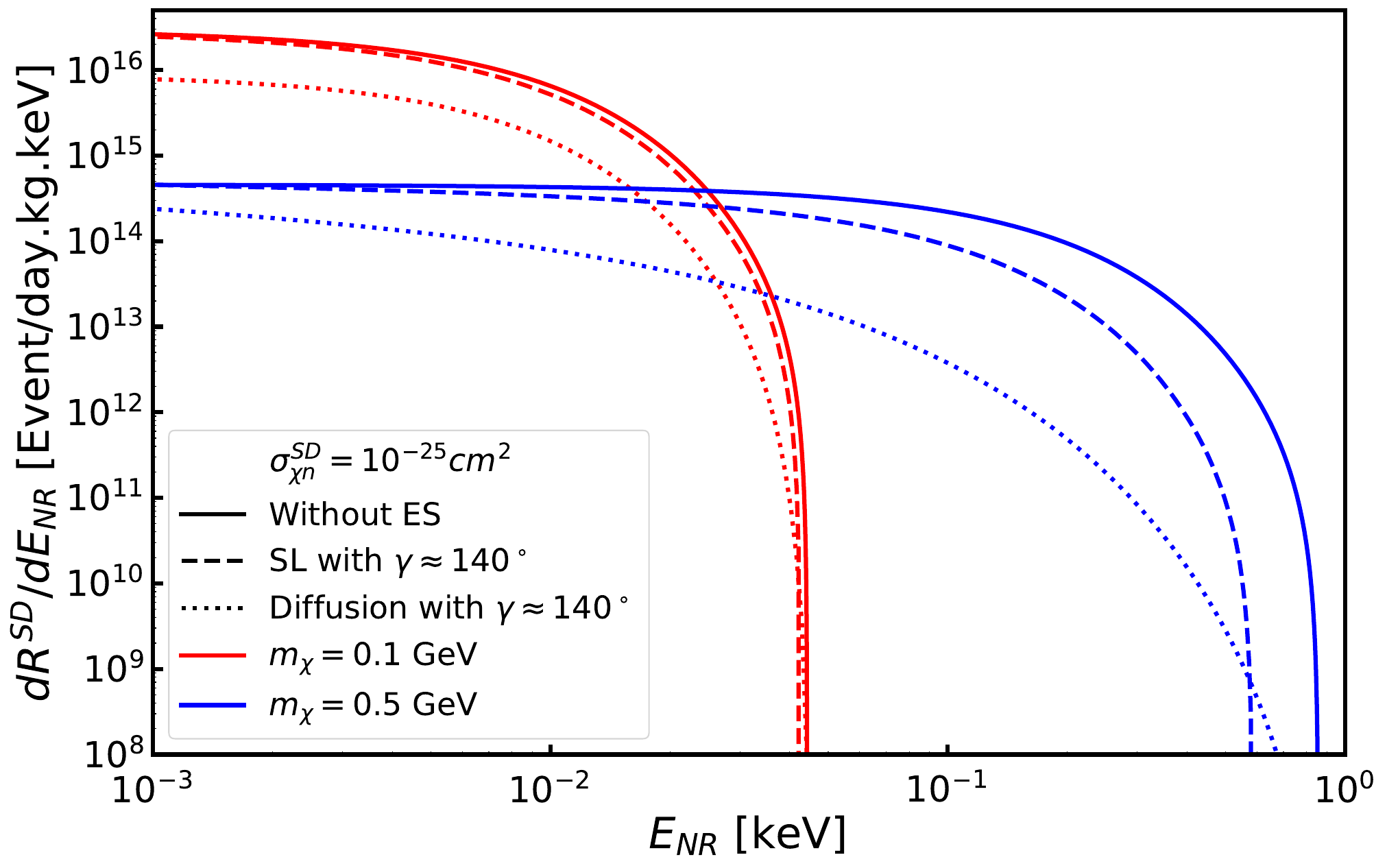}
\includegraphics[width=0.5\textwidth]{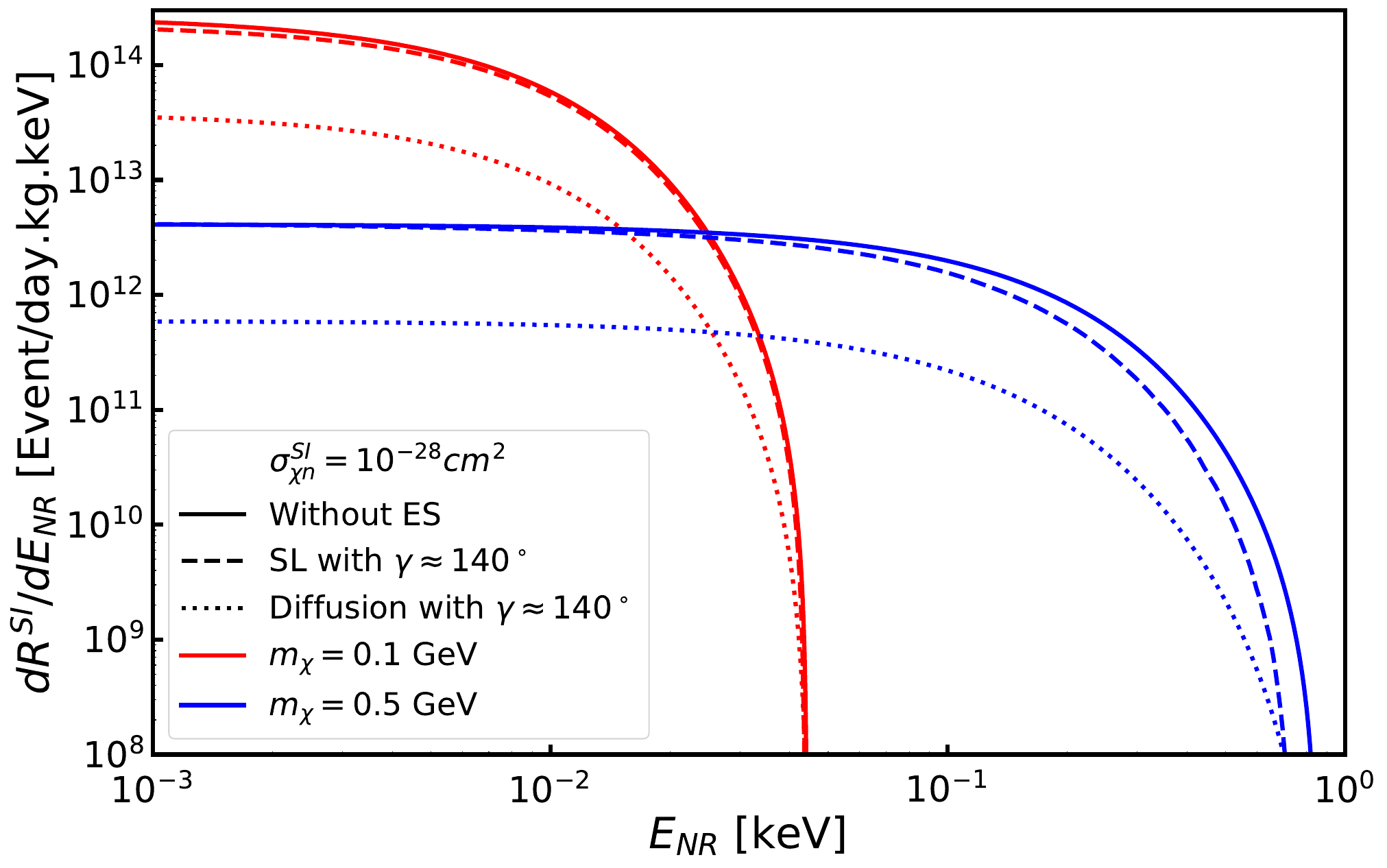}
\caption{Signal spectra for different DM masses $m_\chi=0.1$ GeV and $0.5$ GeV with spin-dependent $\sigma^{\rm SD}_{\rm \chi n}=10^{-25}$cm$^2$ and spin-independent $\sigma^{\rm SI}_{\rm \chi n}=10^{-28}$cm$^2$ interaction cross-sections.  The spectra without the DM stopping effect are shown as solid lines, while those including the stopping effect, using the straight-line (SL) path and the diffusion framework, are represented by dashed and dotted lines, respectively. }
\label{fig:ES-Rate}
\end{figure*}

Then, through an iterative process, we calculate the probability of a DM particle reaching the detector  after $n$ scatterings, as
\begin{equation}
    P_n(z) = \int_0^{\infty} P_{n-1}(z')\frac{1}{2\lambda}\Gamma(0,|z-z'|/\lambda)\mathrm{d}z'\,.
\end{equation}
Both the Gamma function and the probability distribution fall off rapidly as $ z'\to \infty $, so the upper integration limit can be set to a finite value, significantly larger than the atmosphere's height.

We now focus on the energy spectrum and the energy loss of DM after undergoing $n$ scatterings. The energy spectrum is parametrised in terms of the probability of scattering events. The contribution from each scattering depends on the spectrum from the previous one. This iterative process is repeated until the DM particle loses all of its energy. 
\begin{equation}
    P(\Delta E) = \frac{1}{E_{\rm max}}\theta(E_{\rm max} - \Delta E)\,,
\end{equation}
where $\theta$ is the Heaviside step function. In the atmospheric energy loss modelling, we include the contributions of Nitrogen-14 for spin-dependent interactions and both Nitrogen-14 and Oxygen-16 for spin-independent interactions. These contributions are weighted according to their relative abundances and number density, denoted as $n_{\rm N}$:
\begin{align}
\label{eq:diff-P}
&P^{\rm int}(\Delta E)= 
\frac{\sum\limits_{\rm N} \left(n_{\rm N} \sigma_{0}^{\rm int}/E_{{\rm max}, \rm N}\right)   \theta(E_{{\rm max}, \rm N} - \Delta E)}
{\sum\limits_{\rm N}n_{\rm N} \sigma_{0}^{\rm int}}
\end{align}
where int $\equiv$ SD or int $\equiv$ SI, and $\sigma_{0}^{\rm SI}$ and $\sigma_{0}^{\rm SD}$ are given in Eqs.~\eqref{Eq:Sigma0SD} and \eqref{Eq:Sigma0SI}.

Given the energy spectrum from the previous step, $\mathrm{d}N/{\mathrm{d}E}_{n-1}(E)$, the spectrum at the $n$-th iteration, where the DM loses energy $\Delta E$ with a probability $P^{\rm int}(\Delta E)$, is determined by applying a change of variables and integrating over the probability distribution:
\begin{align}
\frac{\mathrm{d}N}{\mathrm{d}E}_n(E) &= \int \left( \frac{E}{E - \Delta E} \right) 
\frac{\mathrm{d}N}{\mathrm{d}E}_{n-1} \left( \frac{E^2}{E - \Delta E} \right) \nonumber \\
&\quad \times P^{\rm int}(\Delta E) \, \mathrm{d}\Delta E \,.
\end{align}
The energy spectrum of DM reaching the detector is expressed as the weighted sum over the spectra after all scatters:
\begin{align}
\frac{\mathrm{d}N}{\mathrm{d}E}_{\text{total}}(z, E) \,=\, \sum_{n=0}^\infty \int_z^\infty P_n(z') \, dz' \, \frac{\mathrm{d}N}{\mathrm{d}E}_n(E) \,.
\end{align}

In practice, this sum can be truncated when additional scatterings contribute a negligible flux above the detection threshold. 

Fig.~\ref{fig:Fv} shows the resulting DM velocity distributions at the detector for two benchmark scenarios. Using the diffusion framework and the straight-line approximation we obtain DM velocity distributions at the detector, which depend on the DM mass and interaction strength. We observe that at moderate cross-sections the straight-line approximation is almost identical to the velocity spectrum obtained under the simplified assumption of our planet being opaque to DM, see Fig.~\ref{fig:shadow} for comparison.  However at larger cross-sections it is intriguing to see that the diffusion calculation results in significantly fewer DM particles reaching the detector. This is the case because a large fraction of DM particles that are lighter than the target are reflected back into space, as also discussed in Ref.~\cite{Leane:2023woh}. In addition, the straight-line approximation at large cross-sections shows spurious features in the velocity spectrum that result from the Earth's geometry, which are not present in the scenario where DM reaches the detector by a random walk.

\section{Upper Limits on Cross-Sections} 
\begin{figure*}[t]
\includegraphics[width=0.51\textwidth]{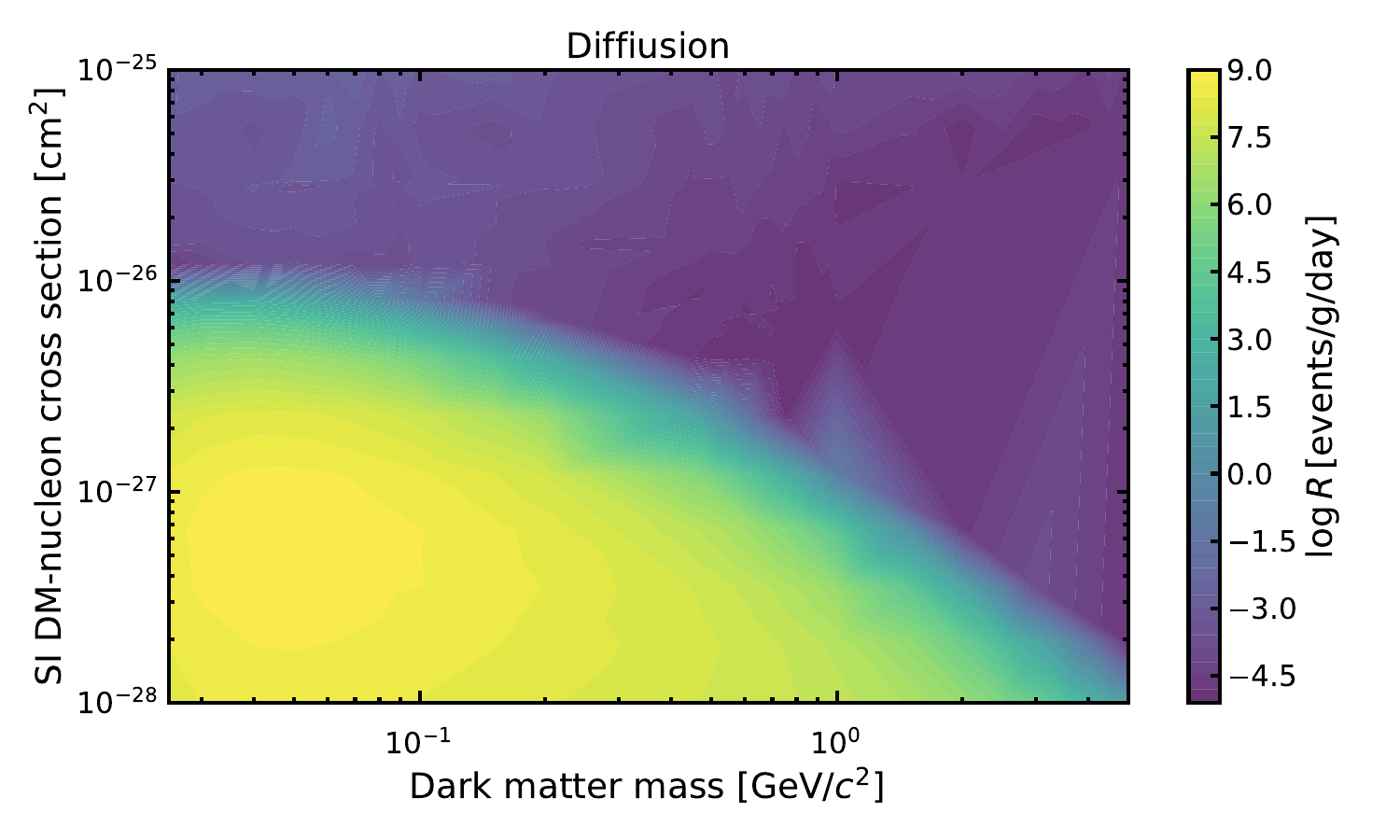}
\includegraphics[width=0.51\textwidth]{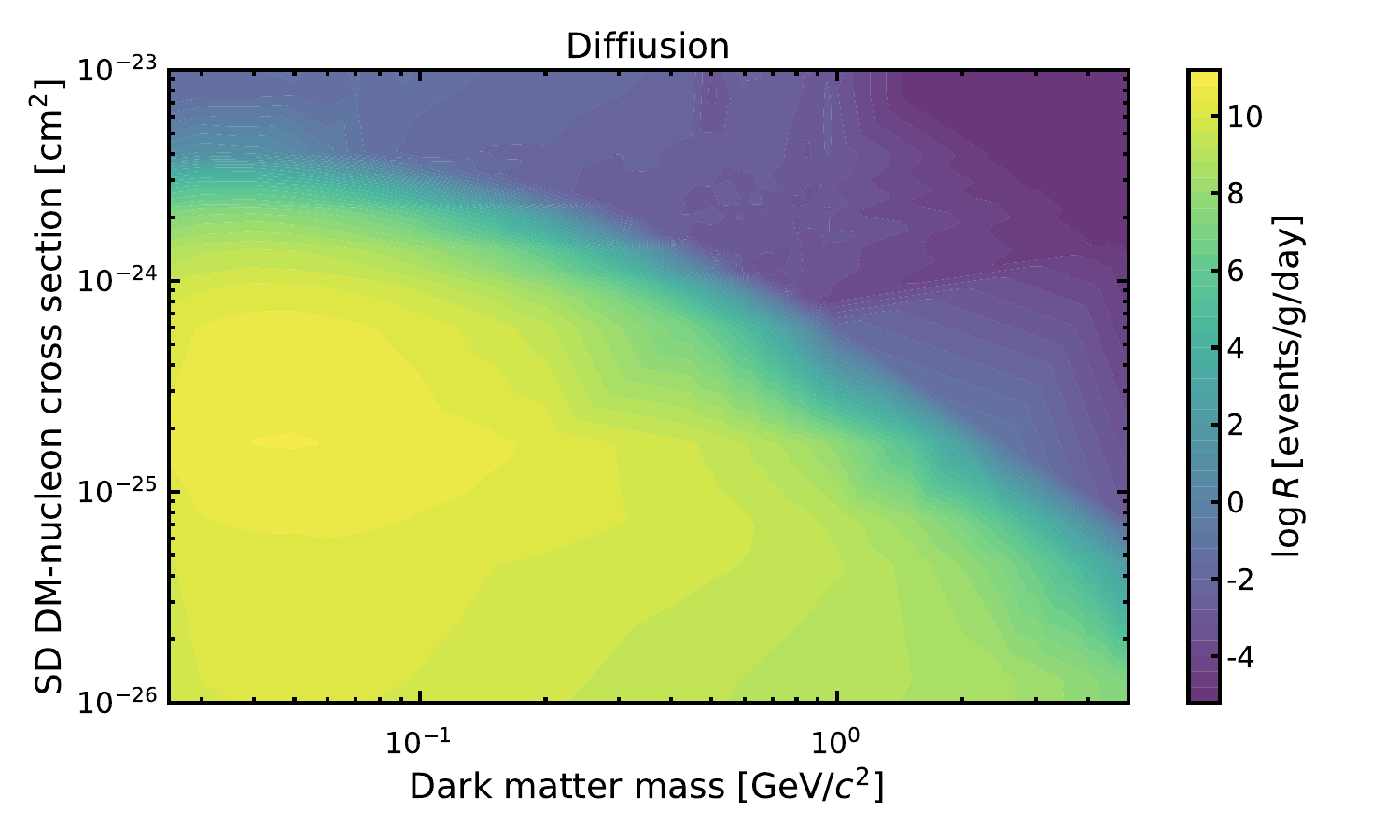}
\includegraphics[width=0.51\textwidth]{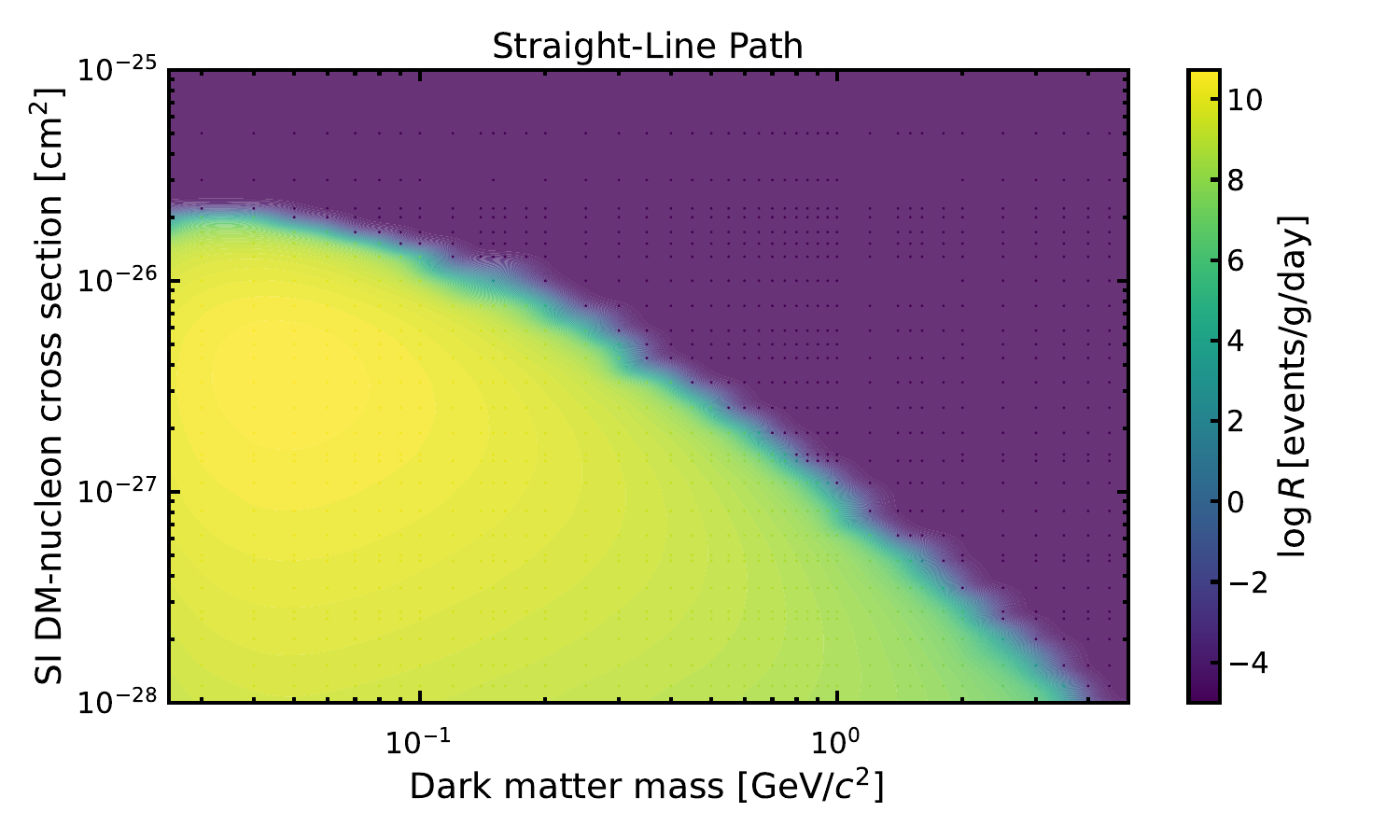}
\includegraphics[width=0.51\textwidth]{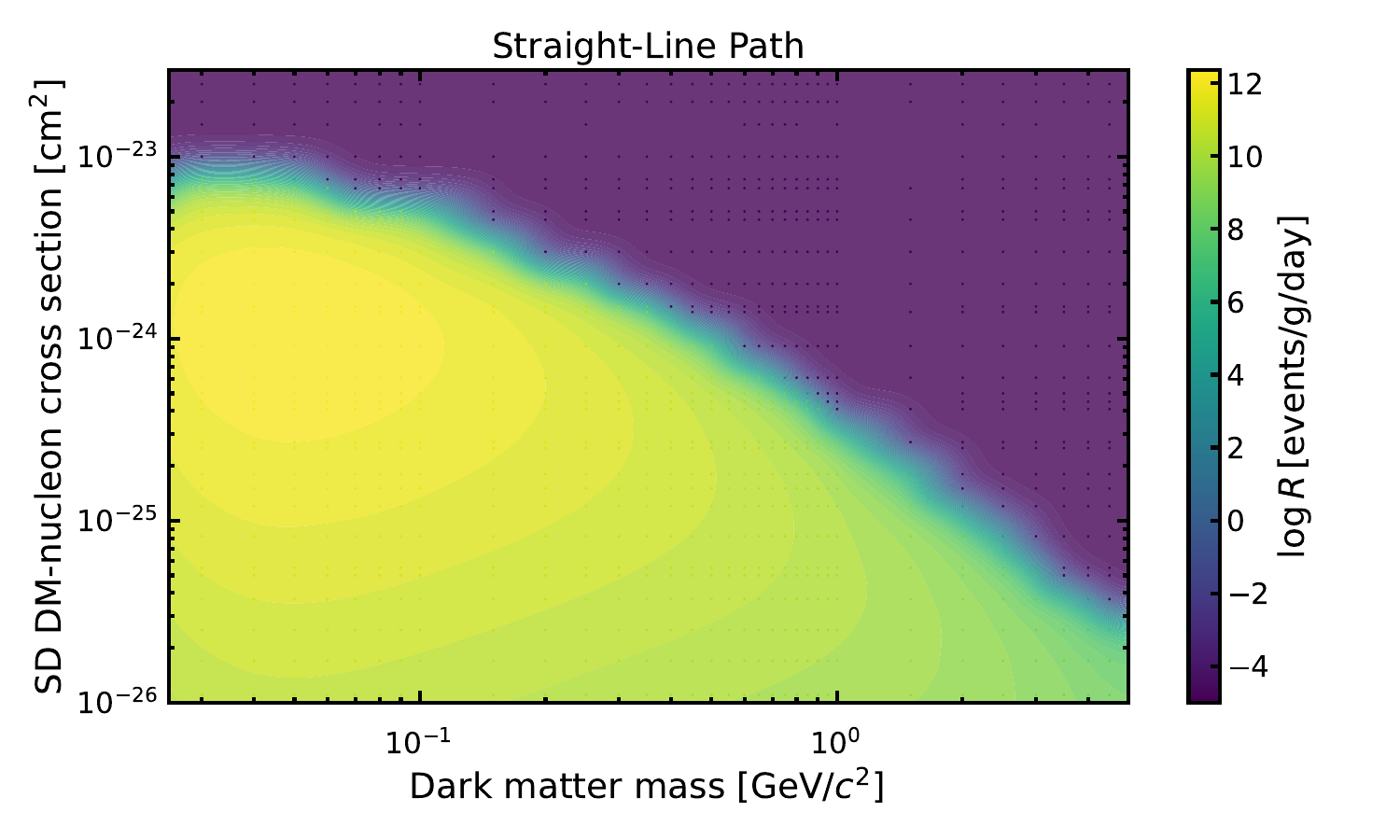}
\caption{Comparison of the predicted DM detection rates for spin-independent and spin-dependent interactions across the extended analytic diffusive and straight-line path models. The top-left panel shows the event rate $\log_{10}R$ for spin-independent interactions under the diffusive model, while the top-right panel displays the corresponding rates for spin-dependent interactions. The bottom-left panel illustrates the spin-independent event rates under the straight-line path model, and the bottom-right panel is the spin-dependent rates for the same model.
}
\label{fig:rate-heatmap}
\end{figure*}

DM scattering impacts both the number density and velocity distribution at the detector location. The differential rate per recoil energy for spin-dependent and spin-independent interactions, as presented in Eqs.~\eqref{drdeSD} and \eqref{drdeSI}, receives corrections proportional to Eq.~\eqref{Eq:dvdDint} for the straight-line path framework, and to Eq.~\eqref{eq:diff-P} for the diffusion framework. 

The impact of these corrections is illustrated in Fig.~\ref{fig:ES-Rate}, which displays the recoil energy spectra in the detector for DM masses of 0.1 GeV and 0.5 GeV, in the straight-line approximation and the diffusive framework described in detail in the preceding section. The differences in these spectra, with and without the Earth-stopping effect, underscore the necessity of accounting for these effects in comprehensive DM studies.   Further, it can be seen that the straight-line approximation and the diffusive framework lead to different results, on the one hand in terms of the total event rate, but also in the recoil energy distribution. This is not surprising as in the case of a random walk the distance covered by a particle scattering isotropically in a medium scales as $d_{\text{Diff}} \sim \lambda \sqrt{n}$, while for a straight-line trajectory we expect $d_{\rm SLP} \sim \lambda \, n$. Thus, a particle undergoing a random walk will scatter significantly more often than one travelling on a straight-line trajectory. Therefore, in the diffusion framework the DM particles are re-scattered to lower recoil energies significantly more often.

The predicted detection rates for spin-independent and spin-dependen interactions, across the diffusion and straight-line path treatment, over a range of DM masses and cross sections are compared in Fig.~\ref{fig:rate-heatmap}. It is evident that the straight-line path approximation overestimates the event rate at the detector, with the discrepancies growing larger at smaller DM masses and larger interaction cross-sections.  

 \begin{figure*}[t]
\includegraphics[width=0.5\textwidth]{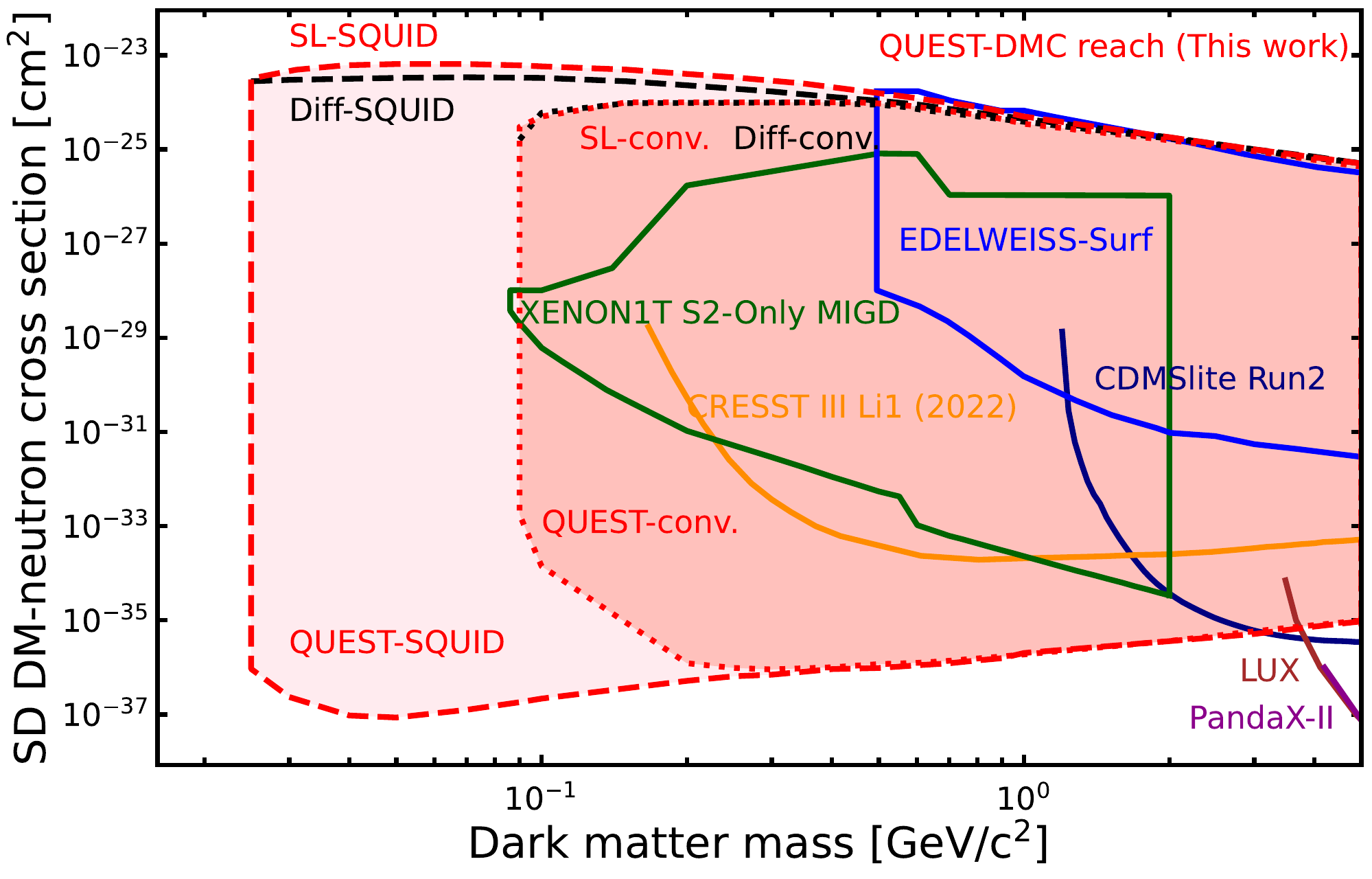}
\includegraphics[width=0.5\textwidth]{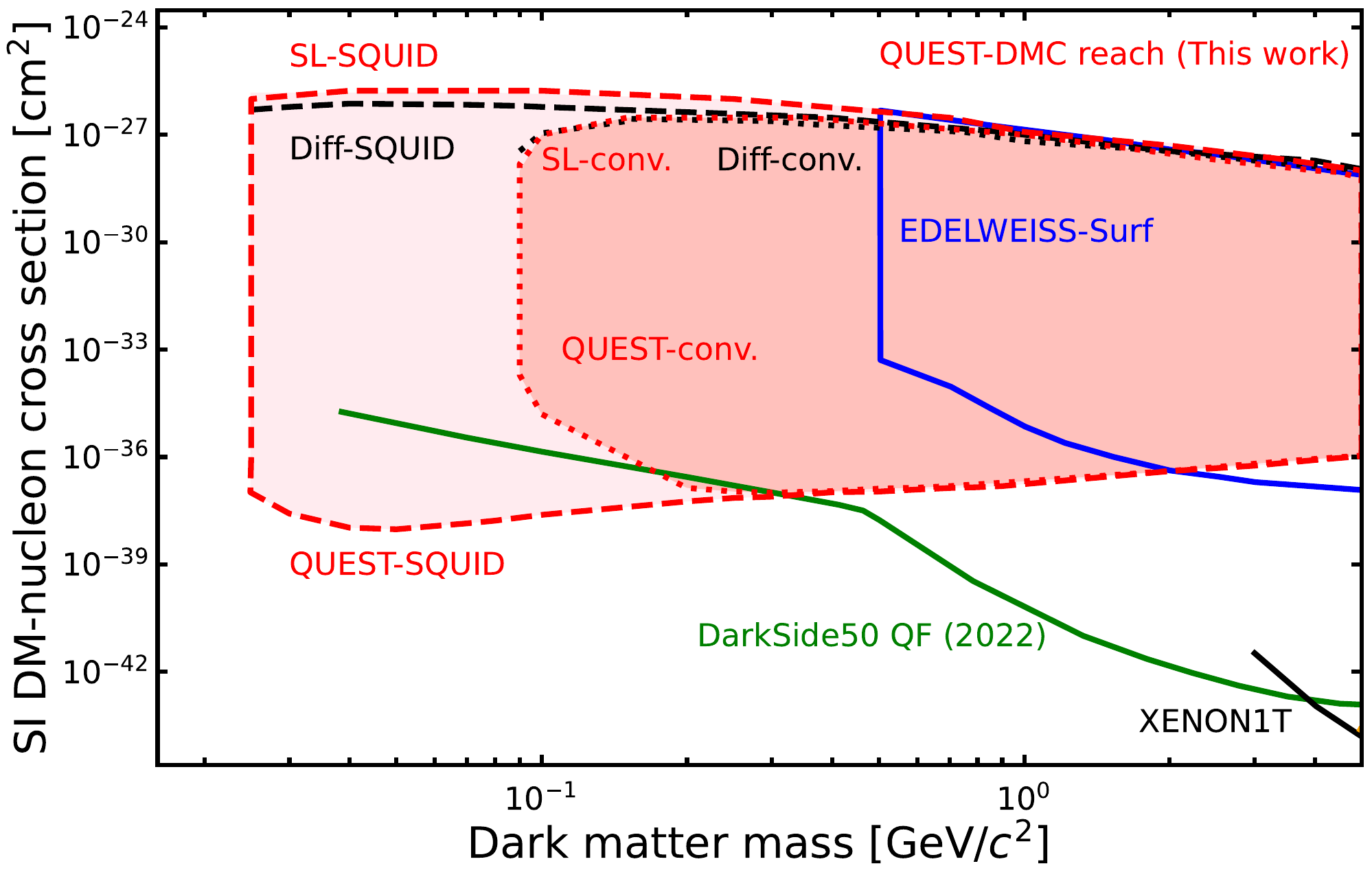}
\caption{The red region shows the range of the projected sensitivity of QUEST-DMC to DM scattering cross-sections for spin-dependent (left panel) and spin-independent (right panel) interactions, using the straight-line and diffusive models described in the text. Solid lines indicate the 90\% confidence level exclusion limits set by other experiments, with closed lines representing experiments where sensitivity ceiling values are available in the literature. Due to its surface location and low energy threshold sensitivity, QUEST-DMC is highly competitive in the unexplored range of large DM scattering cross-sections at low DM masses.}
\label{fig:ES-SD-SI}
\end{figure*}
 
To estimate the projected upper limit sensitivity to the DM interaction cross-section, we incorporate the detector response model into the calculated differential event rate, including DM interactions in the Earth and atmosphere. This response model is described in Ref. \cite{QUEST-DMC:2023nug}.  The detector response modifies the DM signal and the background distributions, which are deployed in a profile likelihood ratio test of the signal plus background relative to the background-only hypothesis. As in~\cite{QUEST-DMC:2023nug}, the sensitivity estimate here assumes an exposure of $4.9$~g.day corresponding to five cells of $0.03$~g Helium-3 operating for 6 months with a 50\% duty cycle, and, an energy threshold of 31 eV based on conventional readout using a cold transformer, whereas using SQUID readout this reaches 0.51 eV.  

Fig.~\ref{fig:ES-SD-SI} presents the upper limits on the projected sensitivity to DM for spin-dependent and spin-independent interactions. 
The spin-dependent DM-neutron scattering sensitivity spans a mass range of $0.025$–$5\,\mathrm{GeV}/c^2$ with SQUID readout and $0.09$–$5\,\mathrm{GeV}/c^2$ with the cold transformer readout. These projections are compared with existing limits from Xenon 1T S2-only MIGD~\cite{XENON:2019zpr}, CRESST III (LiAlO$_2$)~\cite{CRESST_2022}, LUX (Xe)~\cite{LUXSD_2016}, CDMSlite (Ge)~\cite{CDMSLite_2018}, PandaX-II~\cite{PandaX-II:2018woa}, and EDELWEISS~\cite{EDELWEISS:2019vjv}. The upper bound, at 90\% confidence level, from the QUEST-DMC SQUID readout is projected to reach a cross-section sensitivity of $6.5 \times 10^{-24}{\rm cm}^2$ for the straight-line path and $3.3 \times 10^{-24}\,{\rm cm}^2$ for the diffusion framework in a mass range $0.04\!-\!0.07\,{\rm GeV}/c^2$. In contrast, using the cold transformer readout, the upper sensitivity limit is projected to reach $1 \times 10^{-24}\,{\rm cm}^2$ in the mass range of $0.1\!-\!0.55\,{\rm GeV}/c^2$ for both models.

The spin-independent DM-nucleon scattering sensitivity is shown together with existing constraints from DarkSide-50~\cite{DarkSide-50:2022qzh}, XENON1T~\cite{XENON:2018voc}, and EDELWEISS~\cite{EDELWEISS:2019vjv}. The spin-independent projected sensitivity covers the same mass range as the spin-dependent sensitivity for both readout methods. Using the cold transformer the ceiling is projected to reach a cross-section of $3 \times 10^{-27}\,{\rm cm}^2$ at a mass of $0.15\!-\!0.35\,{\rm GeV}/c^2$ for the straight-line path and diffusion frameworks alike. Meanwhile, the SQUID readout constrains the sensitivity at higher values, reaching $1.8 \times 10^{-26}\,{\rm cm}^2$ for the straight-line path and $7.4 \times 10^{-27}\,{\rm cm}^2$ for the diffusion framework in the mass range of $0.04\!-\!0.1\,{\rm GeV}/c^2$.

\section{Conclusion}

This study has explored and expanded the projected sensitivity reach of the QUEST-DMC experiment, focusing on the upper bounds on the cross-sections for sub-GeV DM interactions. By investigating the attenuation effects caused by Earth's atmosphere the analysis provides a refined understanding of how these phenomena shape the detection probabilities of DM particles with varying masses and interaction cross-sections.  
This study establishes a robust framework for predicting QUEST-DMC's sensitivity in future experimental runs.

Central to this analysis is the treatment of Earth shadowing effects, which become critical for large cross-section DM interactions. Two modelling approaches, the straight-line path approximation and the diffusion framework, were compared to assess their impact on detection rates and velocity distributions. The straight-line path approximation, while computationally efficient and useful for initial projections, tends to overestimate the number of particles reaching the detector, particularly for lighter DM masses. In contrast, the diffusion framework, which accounts for isotropic scattering and cumulative energy loss, provides a more accurate representation of DM attenuation dynamics.

The limits for both frameworks are nearly identical when using the conventional readout. However, with lower energy thresholds achievable with SQUID readout, the cross-section coverage extends to higher values, and at certain critical cross-sections and masses below $1~\mathrm{GeV}/c^2$, the deviation between the two frameworks becomes pronounced. The ceiling sensitivity limit on spin-dependent DM-neutron cross-sections is $\sim 3 \times 10^{-24} \mathrm{cm}^2$ using the diffusive framework and reaches approximately two times higher with the straight-line path DM scattering. Similarly, for spin-independent DM-nucleon cross-sections, the limit is $\sim 7.5 \times 10^{-27} \mathrm{cm}^2$ using the diffusive framework and reaches $\sim 1.8 \times 10^{-26} \mathrm{cm}^2$ with the straight-line path approximation, within the mass range of $0.025$–$5~\mathrm{GeV}/c^2$.

The study highlights the enhanced capabilities of QUEST-DMC through the integration of quantum sensing technologies. The projected sensitivity curves demonstrate that the SQUID-based readout offers substantial improvements over conventional methods, and, the projections suggest that QUEST-DMC has the potential to address unexplored regions of parameter space with both its upper and lower DM scattering cross-section reach. 

This study lays a strong foundation for interpreting future experimental results. The refined sensitivity projections underscore the necessity of incorporating Earth shadowing effects into experimental models, ensuring robust and accurate predictions. These findings establish the QUEST-DMC detection concept's potential as a frontrunner in the search for light DM and illustrate the transformative potential of advanced detection techniques in expanding our understanding of the dark sector.

\acknowledgments
This work was funded by UKRI EPSRC and STFC (Grants ST/T006773/1, ST/Y004434/1, EP/P024203/1, EP/W015730/1 and EP/W028417/1), as well as the European Union’s Horizon 2020 Research and Innovation Programme under Grant Agreement no 824109 (European Microkelvin Platform). S.A. acknowledges financial support from the Jenny and Antti Wihuri Foundation. M.D.T acknowledges financial support from the Royal Academy of Engineering (RF/201819/18/2). J.S. acknowledges support from the UK Research and Innovation Future Leader Fellowship~MR/Y018656/1. A.K. acknowledges support from the UK Research and Innovation Future Leader Fellowship MR/Y019032/1.

\bibliographystyle{JHEP}
\bibliography{bib.bib}

\providecommand{\href}[2]{#2}\begingroup\raggedright\begin{thebibliography}{10}

\bibitem{Zurek:2013wia}
K.M.~Zurek, \emph{{Asymmetric Dark Matter: Theories, Signatures, and
  Constraints}},
  \href{https://doi.org/10.1016/j.physrep.2013.12.001}{\emph{Phys. Rept.}
  {\bfseries 537} (2014) 91} [\href{https://arxiv.org/abs/1308.0338}{{\ttfamily
  1308.0338}}].

\bibitem{Barnes:2020vsc}
P.~Barnes, Z.~Johnson, A.~Pierce and B.~Shakya, \emph{{Simple Hidden Sector
  Dark Matter}}, \href{https://doi.org/10.1103/PhysRevD.102.075019}{\emph{Phys.
  Rev. D} {\bfseries 102} (2020) 075019}
  [\href{https://arxiv.org/abs/2003.13744}{{\ttfamily 2003.13744}}].

\bibitem{Hochberg:2014dra}
Y.~Hochberg, E.~Kuflik, T.~Volansky and J.G.~Wacker, \emph{{Mechanism for
  Thermal Relic Dark Matter of Strongly Interacting Massive Particles}},
  \href{https://doi.org/10.1103/PhysRevLett.113.171301}{\emph{Phys. Rev. Lett.}
  {\bfseries 113} (2014) 171301}
  [\href{https://arxiv.org/abs/1402.5143}{{\ttfamily 1402.5143}}].

\bibitem{Pospelov:2008jk}
M.~Pospelov, A.~Ritz and M.B.~Voloshin, \emph{{Bosonic super-WIMPs as keV-scale
  dark matter}}, \href{https://doi.org/10.1103/PhysRevD.78.115012}{\emph{Phys.
  Rev. D} {\bfseries 78} (2008) 115012}
  [\href{https://arxiv.org/abs/0807.3279}{{\ttfamily 0807.3279}}].

\bibitem{Jaeckel:2012mjv}
J.~Jaeckel, \emph{{A force beyond the Standard Model - Status of the quest for
  hidden photons}}, {\emph{Frascati Phys. Ser.} {\bfseries 56} (2012) 172}
  [\href{https://arxiv.org/abs/1303.1821}{{\ttfamily 1303.1821}}].

\bibitem{Hall:2009bx}
L.J.~Hall, K.~Jedamzik, J.~March-Russell and S.M.~West, \emph{{Freeze-In
  Production of FIMP Dark Matter}},
  \href{https://doi.org/10.1007/JHEP03(2010)080}{\emph{JHEP} {\bfseries 03}
  (2010) 080} [\href{https://arxiv.org/abs/0911.1120}{{\ttfamily 0911.1120}}].

\bibitem{QUEST-DMC:2023nug}
{\scshape QUEST-DMC} collaboration, \emph{{QUEST-DMC superfluid $^3$He detector
  for sub-GeV dark matter}},
  \href{https://doi.org/10.1140/epjc/s10052-024-12410-8}{\emph{Eur. Phys. J. C}
  {\bfseries 84} (2024) 248}
  [\href{https://arxiv.org/abs/2310.11304}{{\ttfamily 2310.11304}}].

\bibitem{Autti:2023gxg}
S.~Autti et~al., \emph{{Long nanomechanical resonators with circular
  cross-section}},  \href{https://arxiv.org/abs/2311.02452}{{\ttfamily
  2311.02452}}.

\bibitem{Autti:2024awr}
S.~Autti et~al., \emph{{QUEST-DMC: Background Modelling and Resulting Heat
  Deposit for a Superfluid Helium-3 Bolometer}},
  \href{https://doi.org/10.1007/s10909-024-03142-w}{\emph{J. Low Temp. Phys.}
  {\bfseries 215} (2024) 465}
  [\href{https://arxiv.org/abs/2402.00181}{{\ttfamily 2402.00181}}].

\bibitem{Collar:1992qc}
J.I.~Collar and F.T.~Avignone, \emph{{Diurnal modulation effects in cold dark
  matter experiments}},
  \href{https://doi.org/10.1016/0370-2693(92)90873-3}{\emph{Phys. Lett. B}
  {\bfseries 275} (1992) 181}.

\bibitem{Collar:1993ss}
J.I.~Collar and F.T.~Avignone, III, \emph{{The Effect of elastic scattering in
  the Earth on cold dark matter experiments}},
  \href{https://doi.org/10.1103/PhysRevD.47.5238}{\emph{Phys. Rev. D}
  {\bfseries 47} (1993) 5238}.

\bibitem{Hasenbalg:1997hs}
F.~Hasenbalg et~al., \emph{{Cold dark matter identification: Diurnal modulation
  revisited}}, \href{https://doi.org/10.1103/PhysRevD.55.7350}{\emph{Phys. Rev.
  D} {\bfseries 55} (1997) 7350}.

\bibitem{Kouvaris:2014lpa}
C.~Kouvaris and I.M.~Shoemaker, \emph{{Daily modulation as a smoking gun of
  dark matter with significant stopping rate}},
  \href{https://doi.org/10.1103/PhysRevD.90.095011}{\emph{Phys. Rev. D}
  {\bfseries 90} (2014) 095011}.

\bibitem{Kouvaris:2015laa}
C.~Kouvaris, \emph{{Earth\textquoteright{}s stopping effect in directional dark
  matter detectors}},
  \href{https://doi.org/10.1103/PhysRevD.93.035023}{\emph{Phys. Rev. D}
  {\bfseries 93} (2016) 035023}.

\bibitem{Bernabei:2015nia}
R.~Bernabei et~al., \emph{{Investigating Earth shadowing effect with
  DAMA/LIBRA-phase1}},
  \href{https://doi.org/10.1140/epjc/s10052-015-3473-y}{\emph{Eur. Phys. J. C}
  {\bfseries 75} (2015) 239}.

\bibitem{Kavanagh:2016pyr}
B.J.~Kavanagh, R.~Catena and C.~Kouvaris, \emph{{Signatures of Earth-scattering
  in the direct detection of Dark Matter}},
  \href{https://doi.org/10.1088/1475-7516/2017/01/012}{\emph{JCAP} {\bfseries
  01} (2017) 012}.

\bibitem{Cappiello:2023hza}
C.V.~Cappiello, \emph{{Analytic Approach to Light Dark Matter Propagation}},
  \href{https://doi.org/10.1103/PhysRevLett.130.221001}{\emph{Phys. Rev. Lett.}
  {\bfseries 130} (2023) 221001}
  [\href{https://arxiv.org/abs/2301.07728}{{\ttfamily 2301.07728}}].

\bibitem{Lewin:1995rx}
J.D.~Lewin and P.F.~Smith, \emph{{Review of mathematics, numerical factors, and
  corrections for dark matter experiments based on elastic nuclear recoil}},
  \href{https://doi.org/10.1016/S0927-6505(96)00047-3}{\emph{Astropart. Phys.}
  {\bfseries 6} (1996) 87}.

\bibitem{Savage:2006qr}
C.~Savage, K.~Freese and P.~Gondolo, \emph{{Annual Modulation of Dark Matter in
  the Presence of Streams}},
  \href{https://doi.org/10.1103/PhysRevD.74.043531}{\emph{Phys. Rev. D}
  {\bfseries 74} (2006) 043531}
  [\href{https://arxiv.org/abs/astro-ph/0607121}{{\ttfamily
  astro-ph/0607121}}].

\bibitem{Kopp:2009qt}
J.~Kopp, T.~Schwetz and J.~Zupan, \emph{{Global interpretation of direct Dark
  Matter searches after CDMS-II results}},
  \href{https://doi.org/10.1088/1475-7516/2010/02/014}{\emph{JCAP} {\bfseries
  02} (2010) 014} [\href{https://arxiv.org/abs/0912.4264}{{\ttfamily
  0912.4264}}].

\bibitem{DMStat_2021}
D.~Baxter et~al., \emph{Recommended conventions for reporting results from
  direct dark matter searches},
  \href{https://doi.org/10.1140/epjc/s10052-021-09655-y}{\emph{Eur. Phys. J. C}
  {\bfseries 81} (2021) }.

\bibitem{Kavanagh:2017cru}
B.J.~Kavanagh, \emph{{Earth scattering of superheavy dark matter: Updated
  constraints from detectors old and new}}, {\emph{Phys. Rev. D} {\bfseries 97}
  (2018) 123013} [\href{https://arxiv.org/abs/1712.04901}{{\ttfamily
  1712.04901}}].

\bibitem{Starkman:1990nj}
G.D.~Starkman, A.~Gould, R.~Esmailzadeh and S.~Dimopoulos, \emph{{Opening the
  Window on Strongly Interacting Dark Matter}},
  \href{https://doi.org/10.1103/PhysRevD.41.3594}{\emph{Phys. Rev. D}
  {\bfseries 41} (1990) 3594}.

\bibitem{Davis:2017noy}
J.H.~Davis, \emph{{Probing Sub-GeV Mass Strongly Interacting Dark Matter with a
  Low-Threshold Surface Experiment}},
  \href{https://doi.org/10.1103/PhysRevLett.119.211302}{\emph{Phys. Rev. Lett.}
  {\bfseries 119} (2017) 211302}
  [\href{https://arxiv.org/abs/1708.01484}{{\ttfamily 1708.01484}}].

\bibitem{ISA}
{ISO 2533:1975}, \emph{Standard atmosphere},  1975.

\bibitem{Leane:2023woh}
R.K.~Leane and J.~Smirnov, \emph{{Dark matter capture in celestial objects:
  treatment across kinematic and interaction regimes}},
  \href{https://doi.org/10.1088/1475-7516/2023/12/040}{\emph{JCAP} {\bfseries
  12} (2023) 040} [\href{https://arxiv.org/abs/2309.00669}{{\ttfamily
  2309.00669}}].

\bibitem{XENON:2019zpr}
{\scshape XENON} collaboration, \emph{{Search for Light Dark Matter
  Interactions Enhanced by the Migdal Effect or Bremsstrahlung in XENON1T}},
  \href{https://doi.org/10.1103/PhysRevLett.123.241803}{\emph{Phys. Rev. Lett.}
  {\bfseries 123} (2019) 241803}.

\bibitem{CRESST_2022}
{\scshape CRESST Collaboration} collaboration, \emph{Testing spin-dependent
  dark matter interactions with lithium aluminate targets in {CRESST-III}},
  \href{https://doi.org/10.1103/PhysRevD.106.092008}{\emph{Phys. Rev. D}
  {\bfseries 106} (2022) 092008}.

\bibitem{LUXSD_2016}
D.~Akerib et~al., \emph{Results on the spin-dependent scattering of weakly
  interacting massive particles on nucleons from the {Run 3} data of the {LUX}
  experiment},
  \href{https://doi.org/10.1103/physrevlett.116.161302}{\emph{Phys. Rev. Lett.}
  {\bfseries 116} (2016) }.

\bibitem{CDMSLite_2018}
R.~Agnese et~al., \emph{Low-mass dark matter search with {CDMSlite}},
  \href{https://doi.org/10.1103/physrevd.97.022002}{\emph{Phys. Rev. D}
  {\bfseries 97} (2018) }.

\bibitem{PandaX-II:2018woa}
{\scshape PandaX-II} collaboration, \emph{{PandaX-II Constraints on
  Spin-Dependent WIMP-Nucleon Effective Interactions}},
  \href{https://doi.org/10.1016/j.physletb.2019.02.043}{\emph{Phys. Lett. B}
  {\bfseries 792} (2019) 193}.

\bibitem{EDELWEISS:2019vjv}
{\scshape EDELWEISS} collaboration, \emph{{Searching for low-mass dark matter
  particles with a massive Ge bolometer operated above-ground}},
  \href{https://doi.org/10.1103/PhysRevD.99.082003}{\emph{Phys. Rev. D}
  {\bfseries 99} (2019) 082003}
  [\href{https://arxiv.org/abs/1901.03588}{{\ttfamily 1901.03588}}].

\bibitem{DarkSide-50:2022qzh}
{\scshape DarkSide-50} collaboration, \emph{{Search for low-mass dark matter
  WIMPs with 12~ton-day exposure of DarkSide-50}},
  \href{https://doi.org/10.1103/PhysRevD.107.063001}{\emph{Phys. Rev. D}
  {\bfseries 107} (2023) 063001}.

\bibitem{XENON:2018voc}
{\scshape XENON} collaboration, \emph{{Dark Matter Search Results from a One
  Ton-Year Exposure of XENON1T}},
  \href{https://doi.org/10.1103/PhysRevLett.121.111302}{\emph{Phys. Rev. Lett.}
  {\bfseries 121} (2018) 111302}.

\end{thebibliography}\endgroup
\end{document}